\documentclass[letterpaper,aps,prd,twocolumn,
tightenlines,preprintnumbersnofootinbib,showkeys,superscriptaddress]{revtex4-1}
\pdfoutput=1
\usepackage{XCharter}
\usepackage[T1]{fontenc}
\usepackage{mathptmx}
\usepackage{bm}
\usepackage{bbold}
\usepackage{pifont}
\usepackage{rotating}
\usepackage{fullpage}
\usepackage{amsfonts,shorthand}
\usepackage{amsmath,mathrsfs}
\usepackage{slashed}
\usepackage{siunitx}
\usepackage{amssymb}
\usepackage{graphicx}
\usepackage{epic}
\usepackage{eepic}
\usepackage{epsfig}
\usepackage{latexsym}
\usepackage[dvipsnames]{xcolor}
\usepackage[export]{adjustbox}
\usepackage{float}
\usepackage{multirow, makecell}
\usepackage{hyperref}
\usepackage{enumitem}
\hypersetup{colorlinks=true,citecolor=red,linkcolor=NavyBlue,urlcolor=NavyBlue}
\usepackage[caption=false]{subfig}
 
\usepackage{natbib}
\usepackage{relsize}
\usepackage[left=1.6cm,right=1.6cm,top=1.75cm,bottom=1.84cm]{geometry}
\def\bibfont{\footnotesize}
\begin{document}

\title{Pinning down the leptophobic $Z^\prime$ in leptonic final states with Deep Learning}

\author{Tanumoy Mandal}
\email{tanumoy@iisertvm.ac.in}
\affiliation{Indian Institute of Science Education and Research Thiruvananthapuram, Vithura, Kerala, 695 551, India}

\author{Aniket Masaye}
\affiliation{Indian Institute of Science Education and Research Thiruvananthapuram, Vithura, Kerala, 695 551, India}

\author{Subhadip Mitra}
\email{subhadip.mitra@iiit.ac.in}
\affiliation{Center for Computational Natural Sciences and Bioinformatics, International Institute of Information Technology, Hyderabad 500 032, India}

\author{Cyrin Neeraj}
\email{cyrin.neeraj@research.iiit.ac.in}
\affiliation{Center for Computational Natural Sciences and Bioinformatics, International Institute of Information Technology, Hyderabad 500 032, India}

\author{Naveen Reule}
\email{naveenreule20@iisertvm.ac.in}
\affiliation{Indian Institute of Science Education and Research Thiruvananthapuram, Vithura, Kerala, 695 551, India}

\author{Kalp Shah}
\email{kalp.shah@research.iiit.ac.in}
\affiliation{Center for Computational Natural Sciences and Bioinformatics, International Institute of Information Technology, Hyderabad 500 032, India}

\date{\today}
\preprint{XXXXXX}

\begin{abstract}
\noindent
A leptophobic $Z^\prime$ that does not couple with the Standard Model leptons can evade the stringent bounds from the dilepton-resonance searches. In our earlier paper~[T. Arun \emph{et al.}, Search for the $Z'$ boson decaying to a right-handed neutrino pair in leptophobic $U(1)$ models, \href{http://dx.doi.org/10.1103/PhysRevD.106.095035}{{\it Phys. Rev.} D, {\bf 106} (2022) 095035}; \href{http://arxiv.org/abs/2204.02949}{arXiv:2204.02949}], we presented two gauge anomaly-free $U(1)$ models---one based on the Green-Schwarz (GS) anomaly cancellation mechanism, and the other on a grand unified theory (GUT) framework with gauge kinetic mixing---where a heavy leptophobic $Z'$ is present along with right-handed neutrinos ($N_R$). We pointed out the interesting possibility of a correlated search for $Z'$ and $N_R$ at the LHC through the $pp\to Z'\to N_R N_R$ channel. This channel can probe a part of the $Z'$ parameter space beyond the reach of the standard dijet resonance searches. In this follow-up paper, we analyse the challenging monolepton final state arising from the decays of the $N_R$ pair with Deep Learning. We present the high-luminosity LHC discovery reaches for six different GUT embeddings and a benchmark point in the GS setup. We also update our previous estimates in the dilepton channel with Deep Learning. We identify parameter regions that can be probed with the proposed channel but will remain inaccessible to dijet searches at the HL-LHC.
\end{abstract}

\maketitle

\section{Introduction}
\noindent
Many extensions of the Standard Model (SM), like the grand unified theories
(GUTs)~\cite{Langacker:1980js,London:1986dk,Hewett:1988xc}, contain an electrically neutral colour-singlet heavy  gauge boson, $Z'$. To get such a particle in the spectrum in a bottom-up manner, one can simply add an extra local $U(1)$ symmetry spontaneously broken by a complex scalar field with a TeV-scale vacuum expectation value (vev). The phenomenology of a TeV-scale $Z'$ is well-explored in the literature (see, e.g.,~\cite{Leike:1998wr,Langacker:2008yv}). One of the current programs of the LHC is to look for the signature of $Z^\prime$. However, so far, the focus has been only on the cases where it decays exclusively to the SM particles~\cite{ATLAS:2019fgd,CMS:2019gwf,ATLAS:2019erb,CMS:2021ctt,ATLAS:2016gzy,CMS:2016kgr,ATLAS:2020fry,CMS:2021klu,CMS:2015fhb}. The failure to find the $Z'$ so far motivates us to look for other possibilities with \emph{nonstandard} decay modes of $Z'$, i.e., where it can decay to beyond-the-SM (BSM) particles as well. 

The strongest exclusion limits on the $Z'$ parameter space usually come from the dilepton resonance searches~\cite{ATLAS:2019erb,CMS:2021ctt}. For instance, the current mass exclusion limit on a sequential $Z'$ is about $5$~TeV~\cite{ATLAS:2019erb}. The stringent dilepton exclusion limits can be evaded if the $Z'$ is leptophobic, i.e., it does not couple (or couples very feebly) to the SM leptons. In the leptophobic case, the nonstandard decay modes of the $Z'$ can become important as the branching ratios (BRs) to the new modes can be high. One interesting possibility is when the $Z'$ decays to a pair of right-handed neutrinos (RHNs, $N_R$)~\cite{Ferrari:2002ac,Das:2017flq,Das:2017deo,Cox:2017eme,Chauhan:2021xus,Das:2022rbl}. While appending an additional $U(1)$ gauge symmetry to the SM group, one must ensure the cancellation of all gauge anomalies to maintain gauge invariance and renormalisability of the theory. To cancel the gauge anomalies, one can include RHNs in the particle spectrum~\cite{Ekstedt:2016wyi} or rely on some special mechanism like the  Green-Schwarz (GS) mechanism~\cite{Leontaris:1999wf,Ekstedt:2017tbo}. The minimal GS mechanism does not require RHNs. However, the cancellation of pure gravity anomaly motivates the existence of RHNs. Therefore, many anomaly-free $U(1)$ extensions of the SM contain RHNs since they can also help generate light neutrino masses through various seesaw mechanisms.

\begin{figure}[t!]
\captionsetup[subfigure]{labelformat=empty}
\centering
\includegraphics[width=0.95\columnwidth]{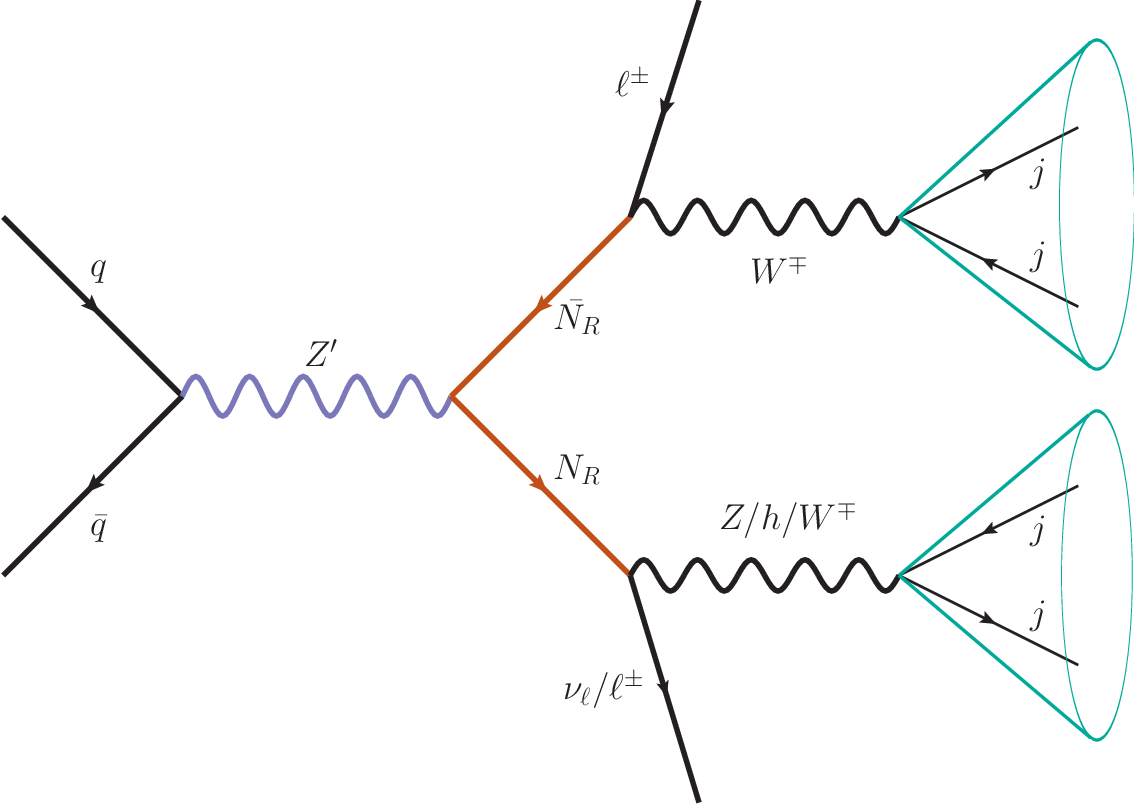}
\caption{Representative Feynman diagram for a leptophobic $Z'$ decaying to two right-handed neutrinos.}
\label{fig:ZprimeFeynDiag}
\end{figure}

In our previous paper~\cite{Arun:2022ecj}, we presented two theoretically well-motivated leptophobic $Z'$ scenarios where the $Z'$ dominantly decays to a pair of RHNs. Being leptophobic, these scenarios can easily bypass the dilepton exclusion limits. At the same time, a large BR of the $Z'\to N_RN_R$ decay and the subsequent decays of $N_R$ to $W\ell$, $Z\nu$, and $H\nu$ final states open up many interesting leptonic final states (see, e.g., Fig.~\ref{fig:ZprimeFeynDiag}). The process $pp\to Z'\to N_RN_R$ is important not only from the $Z'$-search point of view but also for the RHN searches. Since RHNs are SM singlets, producing them at the LHC is difficult as their production cross sections are small, suppressed by the light-heavy neutrino mixing angles. However, they can be copiously produced if they come from the decays of another BSM particle~\cite{Choudhury:2020cpm,Deka:2021koh,ThomasArun:2021rwf,Bhaskar:2023xkm}.

We considered the $pp\to Z'\to N_RN_R$ channel in a leptophobic setup in Ref.~\cite{Arun:2022ecj}. We studied the dilepton channel (which is easy to analyse with a cut-based analysis and shows a good sensitivity) to estimate the HL-LHC reach. Interestingly, we found that a large chunk of the parameter space beyond the reach of $pp\to Z'\to jj$ can be probed through this channel at the HL-LHC. The decay of the RHN pair can lead to multilepton final states. The monolepton mode is complimentary but more challenging because of the difficulty in the background reduction. In this follow-up paper, we estimate the reach in the monolepton final state with a deep neural network (DNN) model.

The RHNs we consider are around the TeV scale, much lighter than the standard type-I seesaw scale $\sim 10^{14}$~GeV. To naturally realise the TeV-scale RHNs, we consider inverse-seesaw mechanism (ISM) for neutrino mass generation~\cite{Mohapatra:1986aw,Mohapatra:1986bd}. As a result, unlike the Majorana-type RHNs in the type-I seesaw, the RHNs here are pseudo-Dirac in nature. If neutrino mass is generated using a type-I seesaw mechanism, because of the Majorana nature of the RHNs, the collider consequences will be different from the ISM. For example, from the $pp\to Z'\to N_RN_R$ process, we get a same-sign dilepton signature, which is not present in the ISM framework. Earlier, in Refs.~\cite{Das:2017kkm,Jana:2019mez,Das:2019pua,Bandyopadhyay:2020djh,Das:2021nqj}, different $U(1)$ extensions using the ISM were considered in various contexts. Also, the future lepton colliders could be the best testing ground for the heavy neutrinos~\cite{Banerjee:2015gca,Chakraborty:2018khw,Barducci:2022hll}.

The rest of the paper is organised as follows. In Sec.~\ref{sec:model}, we review the leptophobic models; in Sec.~\ref{sec:collider}, we present our analysis of the monolepton channel; in Sec.~\ref{sec:monoprospects}, we estimate the HL-LHC prospects of the monolepton channel. We also update our estimate of the dilepton prospects with the DNN in Sec.~\ref{sec:diprospects}. Finally, we conclude in Sec.~\ref{sec:conclu}.

\section{Leptophobic $Z'$ models}\label{sec:model}
\noindent
We look at the two examples of anomaly-free gauge extensions of the SM from Ref.~\cite{Arun:2022ecj}, where a leptophobic $Z'$ with a substantial $Z'\to N_RN_R$ branching is present. In one example, the RHNs cancel the mixed gauge-gravity anomaly while the GS mechanism cancels the rest. In the other example, the leptophobia of $Z'$ arises in a GUT framework~\cite{Babu:1996vt,Lopez:1996ta,Rizzo:1998ut,Leroux:2001fx}.
Below, we briefly review the essential details of these two constructions and elaborate on some aspects for completeness. 

\begin{table}[t!]
\caption{Particle representations and quantum numbers in the GS anomaly-cancellation setup.\label{leptophobic}}
\centering{\renewcommand\baselinestretch{1.5}\selectfont
\begin{tabular*}{\columnwidth}{ l@{\extracolsep{\fill}} rrrr}
\hline
     &  $SU(3)_c$ & $SU(2)_L$ &  $U(1)_Y$ & $U(1)_z$  \\
     \hline\hline
     $Q_L$  & $\mathbf{3}$ & $\mathbf{2}$ &  $1/3$ & $z_q$\\
     
     $u_R$ & $\overline{\mathbf{3}}$ & $\mathbf{1}$ & $4/3$ & $\al z_q$\\
     
     $d_R$ & $\overline{\mathbf{3}}$ & $\mathbf{1}$ & $-2/3$ & $\bt z_q$\\
     
     $\ell_L$ & $\mathbf{1}$ & $\mathbf{2}$ & $-1$ & $0$\\
     
     $e_R$ & $\mathbf{1}$ & $\mathbf{1}$ & $-2$ & $0$\\
     
     $N_R$ & $\mathbf{1}$ & $\mathbf{1}$ & $0$ & $z_N$\\
     
     $H$ & $\mathbf{1}$ & $\mathbf{2}$ & $1$ & 0\\
     
     $\phi$ &  $\mathbf{1}$ & $\mathbf{1}$ & $0$ & $1$\\
    
     $S_L$ &$\mathbf{1}$ & $\mathbf{1}$ & $0$ & $0$\\
    \hline
\end{tabular*}}
\end{table}
\subsection{Leptophobia with GS mechanism}
\label{subsec:LPGS}
\noindent 
In this model, the particle spectrum includes three RHNs (one for every generation) equally charged under the extra $U(1)_z$. Even though the RHNs are not essential for gauge anomaly cancellation with the GS mechanism~\cite{Ekstedt:2017tbo}, the cancellation of the pure gravity anomaly motivates us to introduce them. The light-neutrino masses are generated through the ISM, which requires an extra chiral sterile fermion $S_L$ per generation. 

We show the $U(1)_z$ charges of various particles in Table~\ref{leptophobic}. The SM leptons are uncharged under $U(1)_z$ to make the $Z'$ leptophobic. The nonzero $U(1)_z$ charges of the quarks ensure that the $Z'$ can be produced at the LHC through quark-quark fusion. The $U(1)_z$ charge assignment is generation-independent, and we introduce two free parameters $\al$ and $\bt$ to parametrise the right-handed up and down-type quark charges, respectively ($\al=\bt=-1$ in Ref.~\cite{Arun:2022ecj}). The SM Higgs doublet is chargeless under $U(1)_z$ to minimise the mixing between $Z$ and $Z'$. The scalar $\phi$ with unit $U(1)_z$ charge is the flavon field needed to generate the quark Yukawa interactions as shown in Eq.~\eqref{eq:LagYukflav}. The fermion masses arise after $\phi$ and $H$ get vacuum expectation values.

The charge assignment in Table~\ref{leptophobic} leads to six anomalous triangle diagrams proportional to the traces of the product of the generators:
\begin{enumerate}
    \item $[SU(3)_C]^2[U(1)_z]:\hspace{0.3cm} {\rm tr}[ \{ \mathcal{T}^a,\mathcal{T}^b\} z] = 3z_q(2 - \alpha - \beta)$,
    \item $[SU(2)_L]^2[U(1)_z]:\hspace{0.35cm} {\rm tr}[\{T^a ,T^b\} z] = 6z_q$,
    \item $[U(1)_z]^3:\hspace{1.6cm}  {\rm tr}[z^3] = 3z_q^3(2 - \alpha^3 - \beta^3) - z_N^3$,
    \item $[U(1)_Y]^2[U(1)_z]:\hspace{0.5cm}  {\rm tr}[Y^2z] = \dfrac{2 z_q}{3}(1-8\alpha - 2\beta)$,
    \item $[U(1)_Y][U(1)_z]^2:\hspace{0.5cm}  {\rm tr}[Yz^2] = 2 z_q^2 (1 - 2\alpha^2 + \beta^2)$,
    \item $[R]^2[U(1)_z]:\hspace{1.2cm} {\rm tr}[z] = 3 z_q (2 - \alpha - \beta) - z_N$.
\end{enumerate}
Since gravity becomes prominent at high scales, it is natural to assume that the mixed gauge-gravity anomaly vanishes at low energy without the GS mechanism. Therefore, we set 
\begin{align}
3 z_q (2 - \alpha - \beta) - z_N = 0\    
\end{align}
by hand. This gives us a relation between the $U(1)_z$ charges of the quarks and RHNs. 

To cancel the other anomalies with the 
GS mechanism~\cite{Anastasopoulos:2006cz,Anastasopoulos:2008jt}, we add new gauge-dependent terms in the Lagrangian with carefully chosen coefficients. The Peccei-Quinn (PQ) terms for the mixed gauge anomalies and the $[U(1)_z]^3$ anomaly can be written as
\begin{align}\label{pqterm}
\mathcal{L}_\text{PQ} = &\dfrac{1}{96\pi^2}
\lt(\dfrac{\Theta}{M}\rt)\varepsilon_{\mu\nu\rho\sigma}  \Big[g_z^2 \mc C_{zzz} F_z^{\mu\nu}
F_z^{\rho\sigma}+g_z g'\mc C_{zzy}  F_z^{\mu\nu} F_Y^{\rho\sigma}\nn\\
&+g'^2\mc C_{zyy}  F_Y^{\mu\nu} F_Y^{\rho\sigma}
+ g^2\mc D_{2}  \mathrm{tr}\left(F_W^{\mu\nu}
F_W^{\rho\sigma}\right)\nn\\
&+g_S^2\mc D_{3}  \mathrm{tr}\left(F_S^{\mu\nu}
F_S^{\rho\sigma}\right)\Big]
\end{align}
Under the new $U(1)_z$, the pseudoscalar (Goldstone) axion $\Theta$ transforms as $\Theta \rightarrow \Theta + Mg_z\theta_z$ and the new gauge field transforms as $B^\mu_z \rightarrow B^\mu_z -\partial^\mu \theta_z$. Here, $M$ is the scale at which $U(1)_z$ breaks through the St\"{u}ckelberg mechanism, ${F_z,\ F_Y,\ F_W,\ F_S}$ are the field strengths and ${g_z,\ g',\ g,\ g_S}$ are the coupling constants associated with the ${U(1)_z,\ U(1)_Y,\ SU(2)_L,\ SU(3)_c}$ gauge groups, respectively.
The generalised Chern-Simons (GCS) terms for other mixed anomalies are given as
\begin{align}\label{gcsterm}
\mathcal{L}_\text{GCS}=&\frac{1}{48 \pi^2}\varepsilon_{\mu\nu\rho\sigma}\Big[\vphantom{\Omega_S^{\nu\rho\sigma}} g'^2 g_z \mc E_{zyy} B_Y^\mu B_z^\nu F_Y^{\rho\sigma}+g' g_z^2 \mc E_{zzy} B_Y^\mu B_z^\nu F_z^{\rho\sigma}\nn\\
&+ g^2 g_z \mc K_{2} B_z^\mu  \Omega_W^{\nu\rho\sigma}+g_S^2 g_z \mc K_{3} B_z^\mu\Omega_S^{\nu\rho\sigma}  \Big]\ 
\end{align}
with
\begin{align*}
\Omega_{S,W}^{\nu\rho\sigma} = &\frac{1}{3}\mathrm{tr}\Big[A_{S,W}^{\nu}\left(F_{S,W}^{\rho\sigma}-[A_{S,W}^{\rho},A_{S,W}^{\sigma}]\right)\nn\\
&+\left(\mathrm{cyclic~permutations~of~} \rho, \sigma, v\right)\Big]
\end{align*}
and $A_X = \left[ G,W \right]$. The coefficients $\mc C$, $\mc D$, $\mc E$, and $\mc K$ can be solved in terms of the $U(1)_z$ charges of the fermions as
\begin{align}
\mc{C}_{zzz} &= -3z_q^3\left(2 - \alpha^3 - \beta^3\right) + z_N^3, \nn\\
\mc{C}_{zyy} &= 2z_q \left(2\beta - 8\alpha -1\right)= \frac{3}{2} \mc{E}_{zyy}, \nn\\
\mc{C}_{zzy} &= 6z_q\left(2\alpha^2 - \beta^2 -1\right) = 3E_{zzy}, \\
\mc{D}_2 &= -18z_q = -\frac{3}{2} \mc{K}_2, \nn\\
\mc{D}_3 &= 9z_q \left(\alpha + \beta -2\right) = -\frac{3}{2} \mc{K}_3,\nn
\end{align}
to cancel the anomalies. The $U(1)_z$ invariant Yukawa interactions are written as
\begin{equation}
\label{eq:LagYukflav}
    \mathcal{L}_Y \supset -\lambda_u \left( \frac{\phi}{\Lambda}\right)^{2z_q} \overline{Q}_L \widetilde{H} u_R - \lambda_d \left(\frac{\phi}{\Lambda}\right)^{2z_q} \overline{Q}_L H d_R + h.c.,
\end{equation}
where $\widetilde{H} = i\sigma_2 H^*$, and $\lambda_u$ and $\lambda_d$ are the Yukawa couplings. After the spontaneous breaking of the $U(1)_z$ symmetry, $\phi$ acquires a vev, $v_\phi = \langle \phi\rangle$, leading to the SM Yukawa terms. We set $z_q=1/2$ for our analysis. This fixes $z_N=6$ for $\al=\bt=-1$.

\subsubsection{Neutrino mass generation}
\noindent
The neutrino masses are generated by the ISM through the following higher-dimensional operators,
\begin{align}
\mathcal{L}_Y \supset&\  -\lambda^\nu_{ij} \left( \frac{\phi^\dagger}{\Lambda} \right)^{z_N} \overline{L}^i_L \widetilde{H} N^j_R - {M_R}_{ij} \left( \frac{\phi}{\Lambda} \right)^{z_N} \overline{N}_R^i S_L^j\nn\\
&\ - \frac{\mu}{2} \overline{S}^c_L S_L + h.c.
\end{align}
We get the mass matrix for the neutrinos after the sequential breaking of the $U(1)_z$ and electroweak symmetries in the $(\nu _L^c\ N_R\ S_L^c )^T$ basis as,
\begin{equation}\label{ism}
\mathbf{M_\nu} = \begin{pmatrix}
    0& \mathbf{m_D}& 0\\
    \mathbf{m_D^T}&0&\mathbf{M_R}\\
    0&\mathbf{M_R^T}&\mathbf{\mu}\\
\end{pmatrix},\end{equation} \\
where $\mathbf{m_D} = \mathbf{\lambda^\nu} v_h $, $\mathbf{\mu} $, and $\mathbf{M_R} $ are  $3\times3$ mass matrices in the generation space in general. The parameter $\mathbf{\mu}$ is typically associated with lepton number violation and restores the lepton number symmetry in the $\mathbf{\mu} \rightarrow 0$ limit. 
After diagonalising the matrix in Eq. \eqref{ism}, we get the active (light) neutrino mass matrix as~\cite{Hettmansperger:2011bt},
\begin{equation}
\mathbf{m_\nu} = \mathbf{m_D} \mathbf{M^{-1}_R} \mathbf{\mu} (\mathbf{M_R^T})^{-1}\mathbf{m_D}  + \mathcal{O}\left(\mu \frac{\mathbf{m_d^4}}{\mathbf{M^4_R}}\right).
\end{equation} 
There is a double suppression of the light neutrino masses from the large $\mathbf{M_R}$ (taken around the TeV scale) and the small $\mathbf{\mu}$.

\begin{table*}[t]
\caption{\label{tab:chargeGUT} \small Quantum numbers ($\overline{Q}_\psi \equiv 2\sqrt{6}Q_\psi$, $\overline{Q}_\chi \equiv 2\sqrt{10}Q_\chi$) of the particle contained in the $\mathbf{27}$ representation of $E_6$ for the six possible embeddings---the first one is called the standard embedding. Here, $\mathcal{G}_{\text{SM}}\equiv SU(3)_c\times SU(2)_L\times U(1)_Y$. The $\tan\theta$ and $\delta$ values needed to get the \emph{leptophobic} $Z'$ boson in each embedding are shown in the bottom rows.  }
\centering{\renewcommand\baselinestretch{1.5}\selectfont
\begin{tabular*}{\textwidth}{ l@{\extracolsep{\fill}} lrrrrrrrrrrrr}
\hline
\multirow{2}{*}{Particle} & \multirow{2}{*}{$\mc G_{\rm SM}$\quad} & \multicolumn{2}{c}{I} & \multicolumn{2}{c}{II} & \multicolumn{2}{c}{III} & \multicolumn{2}{c}{IV} & \multicolumn{2}{c}{V} & \multicolumn{2}{c}{VI}\\
\cline{3-4}\cline{5-6}\cline{7-8}\cline{9-10}\cline{11-12}\cline{13-14}
& & $\overline{Q}_\psi$ & $\overline{Q}_\chi$ & $\overline{Q}_\psi$& $\overline{Q}_\chi$& $\overline{Q}_\psi$& $\overline{Q}_\chi$& $\overline{Q}_\psi$& $\overline{Q}_\chi$& $\overline{Q}_\psi$& $\overline{Q}_\chi$& $\overline{Q}_\psi$& \multicolumn{1}{c}{$\overline{Q}_\chi$}\\
\hline \hline
$Q = (u,d)^T$ & $(\textbf{3},\textbf{2})_{1/3}$ & $1$ & $-1$ & $1$ & $-1$ & $1$ & $-1$ & $1$ & $-1$ & $1$ & $-1$ & $1$ & $-1$\\
$L = (\nu,e)^T$ & $(\textbf{1},\textbf{2})_{-1}$ & $1$ & $3$ & $-2$ & $-2$ & $1$ & $3$ & $-2$ & $-2$ & $1$ & $3$ & $-2$ & $-2$\\
$u^c$ & $(\overline{\textbf{3}},\textbf{1})_{-4/3}$ & $1$ & $-1$ & $1$ & $-1$ & $1$ & $3$ & $1$ & $3$ & $-2$ & $-2$ & $-2$ & $-2$\\
$d^c$ & $(\overline{\textbf{3}},\textbf{1})_{2/3}$ & $1$ & $3$ & $1$ & $3$ & $1$ & $-1$ & $1$ & $-1$ & $1$ & $-1$ & $1$ & $-1$\\
$e^c$ & $(\textbf{1},\textbf{1})_{2}$ & $1$ & $-1$ & $1$ & $-1$ & $1$ & $-5$ & $1$ & $-5$ & $4$ & $0$ & $4$ & $0$\\
$\nu^c$ & $(\textbf{1},\textbf{1})_0$ & $1$ & $-5$ & $1$ & $-5$ & $1$ & $-1$ & $1$ & $-1$ & $1$ & $-1$ & $1$ & $-1$\\
$\mathfrak L = (N , E)^T$ & $(\textbf{1},\textbf{2})_{-1}$ & $-2$ & $-2$ & $1$ & $3$ & $-2$ & $-2$ & $1$ & $3$ & $-2$ & $-2$ & $1$ & $3$\\
$\mathfrak L^c = (N^c , E^c)^T$ & $(\textbf{1},\textbf{2})_1$ & $-2$ & $2$ & $-2$ & $2$ & $-2$ & $2$ & $-2$ & $2$ & $-2$ & $2$ & $-2$ & $2$\\
$\mathfrak D$ & $(\textbf{3},\textbf{1})_{-2/3}$ & $-2$ & $2$ & $-2$ & $2$ & $-2$ & $2$ & $-2$ & $2$ & $-2$ & $2$ & $-2$ & $2$\\
$\mathfrak D^c$ & $(\overline{\textbf{3}},\textbf{1})_{2/3}$ & $-2$ & $-2$ & $-2$ & $-2$ & $-2$ & $-2$ & $-2$ & $-2$ & $1$ & $3$ & $1$ & $3$\\
$\widetilde\nu^c$ & $(\textbf{1},\textbf{1})_0$ & $4$ & $0$ & $4$ & $0$ & $4$ & $0$ & $4$ & $0$ & $1$ & $-5$ & $1$ & $-5$\\
\hline
\multicolumn{2}{c}{$\tan\theta$} & \multicolumn{2}{c}{$\sqrt{3/5}$} & \multicolumn{2}{c}{$\sqrt{3/5}$} & \multicolumn{2}{c}{$\sqrt{15}$} & \multicolumn{2}{c}{$\sqrt{5/27}$} & \multicolumn{2}{c}{$\sqrt{5/3}$} & \multicolumn{2}{c}{$0$}\\
\multicolumn{2}{c}{$\delta$} & \multicolumn{2}{c}{$ -1/3$} & \multicolumn{2}{c}{$-1/3$} & \multicolumn{2}{c}{$-\sqrt{10}/3$} & \multicolumn{2}{c}{$-\sqrt{5/12}$} & \multicolumn{2}{c}{$-\sqrt{5/12}$} & \multicolumn{2}{c}{$-\sqrt{10}/3$}\\
\hline
\end{tabular*}}
\end{table*}

\subsection{Leptophobia in $E_6$ GUT models}
\noindent 
A leptophobic $Z^\prime$ can be realised in some $E_6$ GUT models through gauge-boson kinetic mixing~\cite{Babu:1996vt} even though the fermion couplings to the gauge bosons are not arbitrary free parameters. The $U(1)_z $ group in such models can come from a symmetry-breaking chain, like,
\begin{equation}
\label{eq:symchain}
\begin{split}    
E_6 & \rightarrow SO(10) \times U(1)_\chi \\
& \rightarrow SU(5) \times U(1)_\chi \times U(1)_\psi \\
& \rightarrow SU(2)_L \times U(1)_Y \times U(1)_z \\
& \rightarrow SU(2)_L \times U(1)_Y.
\end{split}
\end{equation}
The fundamental representation of $E_6$ is \textbf{27} dimensional, containing all the SM fermions along with two colour-neutral isosinglets $\nu^c$ and  $\widetilde{\nu}^c$, a colour-singlet isodoublet $\mathfrak L = (N\ E)^T$ and its conjugate $\mathfrak L^c$, and a colour-triplet isosinglet $\mathfrak D$ and its antiparticle $\mathfrak D^c$. The RHN can be identified with either $\nu^c$ or $\widetilde{\nu}^c$. Under $SO(10)$, the
\textbf{27} dimensional representation breaks into a \textbf{16}, a \textbf{10} and an \textbf{1}. These further break into representations of $SU(5)$. In the \emph{standard} embedding (I in Table~\ref{tab:chargeGUT}), we put all the SM fermions and the RHN into the \textbf{16}, and the new fermions in the \textbf{10} and the \textbf{1}. There are five other ways one can embed these fermions, leading to a total of six different embeddings~\cite{Leroux:2001fx}. In Table~\ref{tab:chargeGUT}, we show the new charges of all the fermions for the six different embeddings.

From the chain in Eq.~\eqref{eq:symchain}, the $U(1)_z$ charge $Q_z$ can be expressed as $Q_z = Q_\psi \cos\theta - Q_\chi \sin \theta$ where $\theta$ is the $E_6$ mixing angle. However, there is no solution for $\theta$ where the lepton-$Z'$ couplings vanish.  Therefore, we look at the $U(1)_Y\leftrightarrow U(1)_z$ gauge-kinetic mixing terms allowed by the gauge invariance,
\begin{equation}\label{kinetic_mixing}
\mathcal{L}_{kin} \supset -\frac{1}{4} \widetilde{F}_Y^{\mu \nu} \widetilde{F}_{Y\, \mu \nu} - \frac{1}{4} \widetilde{F}_z^{\mu \nu} \widetilde{F}_{z\,\mu \nu} - \frac{\sin \chi}{2} \widetilde{F}_{Y\,\mu \nu} \widetilde{F}^{\mu \nu}_z.
\end{equation}
The kinetic mixing, parametrised by $\sin\chi$, can be removed by the following transformation~\cite{Rizzo:1998ut}:
\begin{equation}
\label{unitary_transformation}
\tilde{B}^{\mu}_Y = B_Y^{\mu} - \tan \chi\ B_z^{\mu}, \quad \tilde{B}_z^{\mu} = \frac{B_z^{\mu}}{\cos \chi}.
\end{equation}
However, this cancellation is valid only at a particular scale, implying that the mixing term can regenerate at higher orders. This makes the couplings energy-dependent. Hence, they must be evaluated at the TeV scale before relating to the experiments. After the above rotation, the couplings of the abelian gauge bosons with the fermions are given as~\cite{Rizzo:1998ut},
\begin{align}
\mathcal{L}_{int} \supset -\overline{\psi} \gamma_\mu \left[ g' \dfrac{Y}{2}B^\mu_Y + g_z\left( Q_z+\sqrt{\frac{3}{5}}\delta \dfrac{Y}{2}\right)B_{z}^\mu\right]\psi,    
\end{align}
where $\delta = -\Tilde{g}_Y\sin\chi/\Tilde{g}_z$. Now, the fermion couplings are functions of two free parameters $\theta$ and $\delta$; we can, in general, make the couplings of any two fields vanish. For example, for the \emph{standard} embedding (Table~\ref{tab:chargeGUT}), leptophobia corresponds to $\theta = \tan^{-1}\sqrt{3/5} $ and $\delta = -1/3$~\cite{Leroux:2001fx}.  Different combinations of $\theta$ and $\delta$ work in different embeddings, as shown in Table~\ref{tab:chargeGUT}.

\begin{figure*}
\centering
\captionsetup[subfigure]{labelformat=empty}
\subfloat[(a)]{\includegraphics[width=0.32\textwidth]{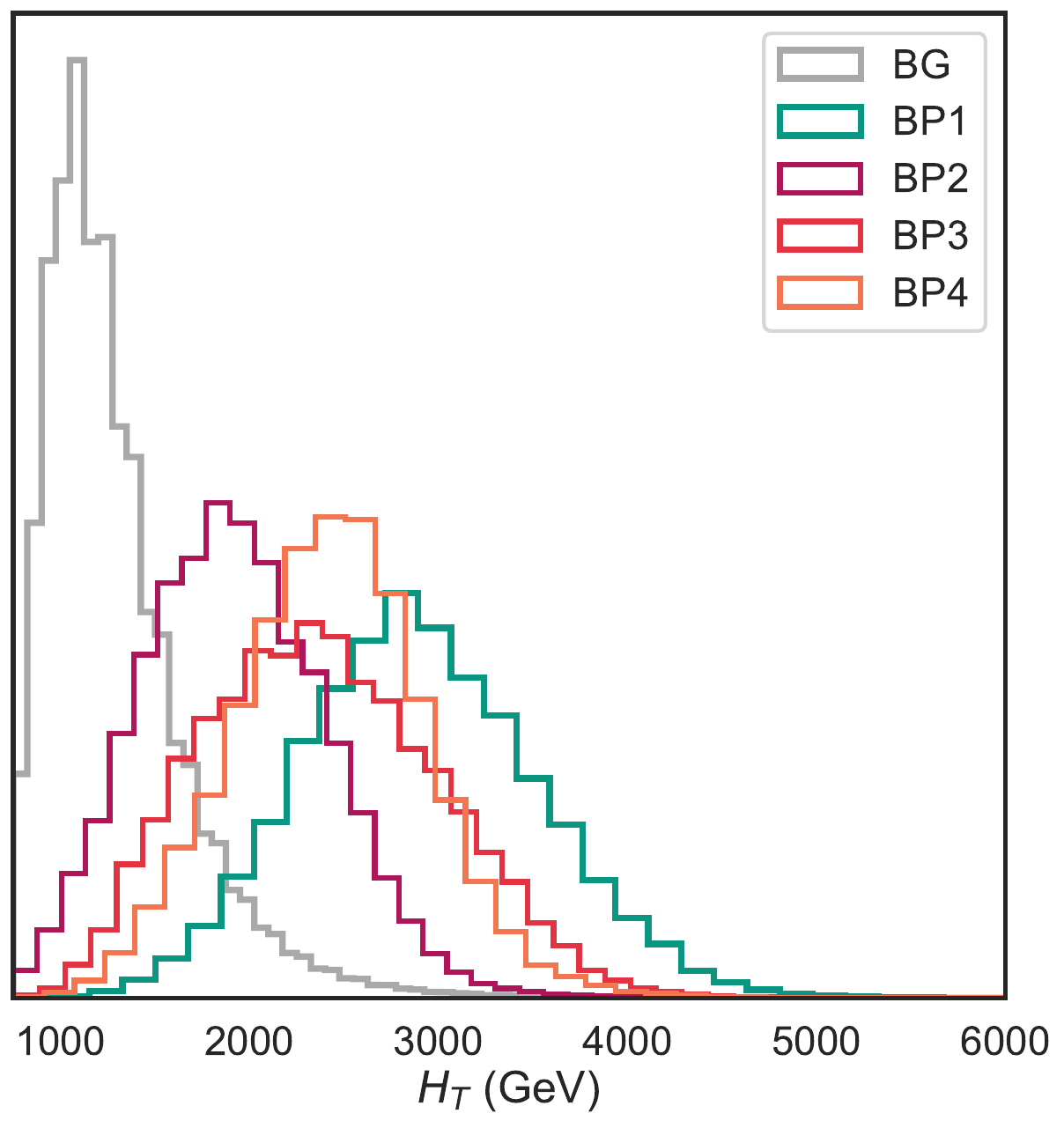}\label{fig:featureA}}\quad
\subfloat[(b)]{\includegraphics[width=0.32\textwidth]{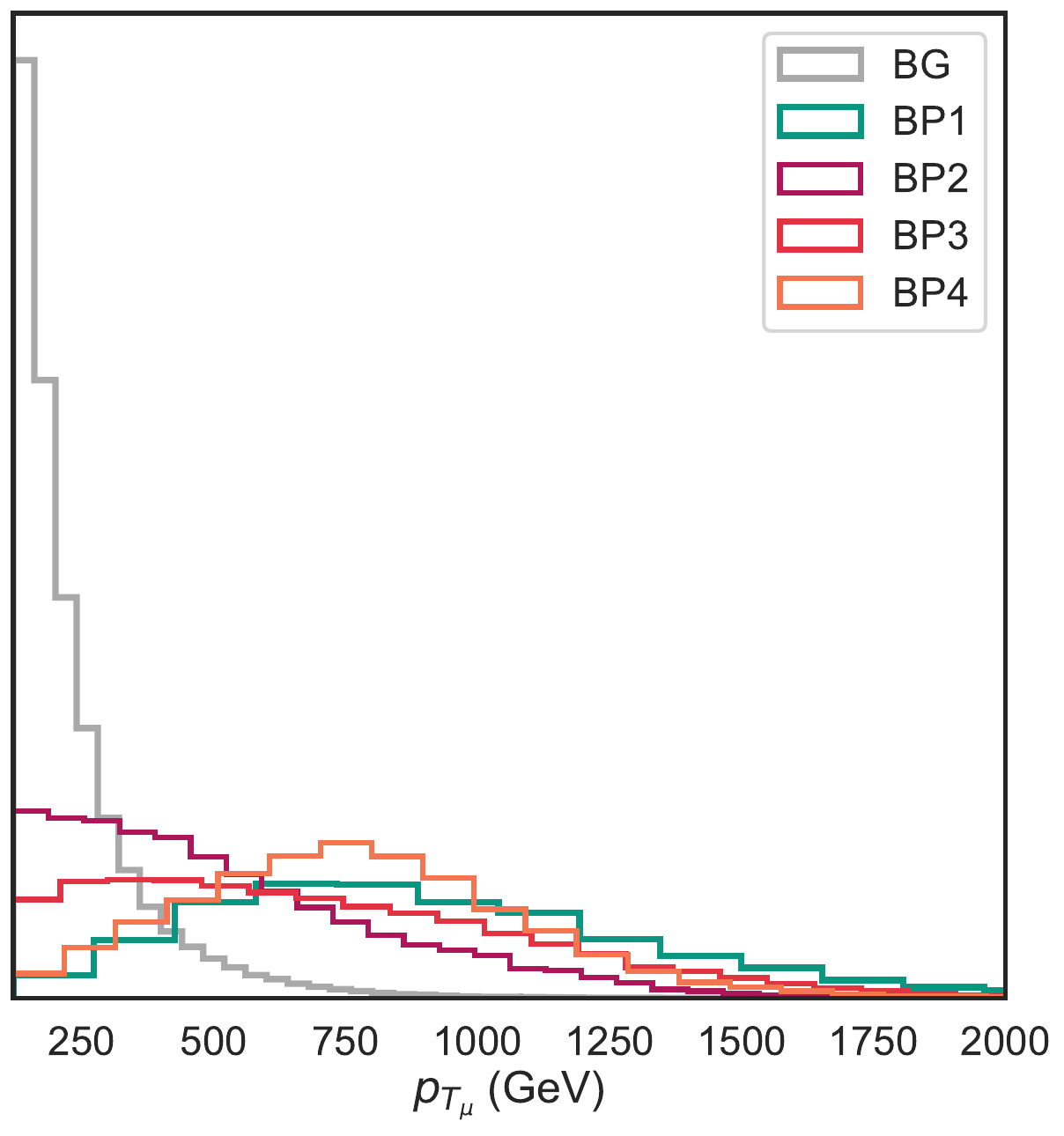}\label{fig:featureB}}\quad
\subfloat[(c)]{\includegraphics[width=0.32\textwidth]{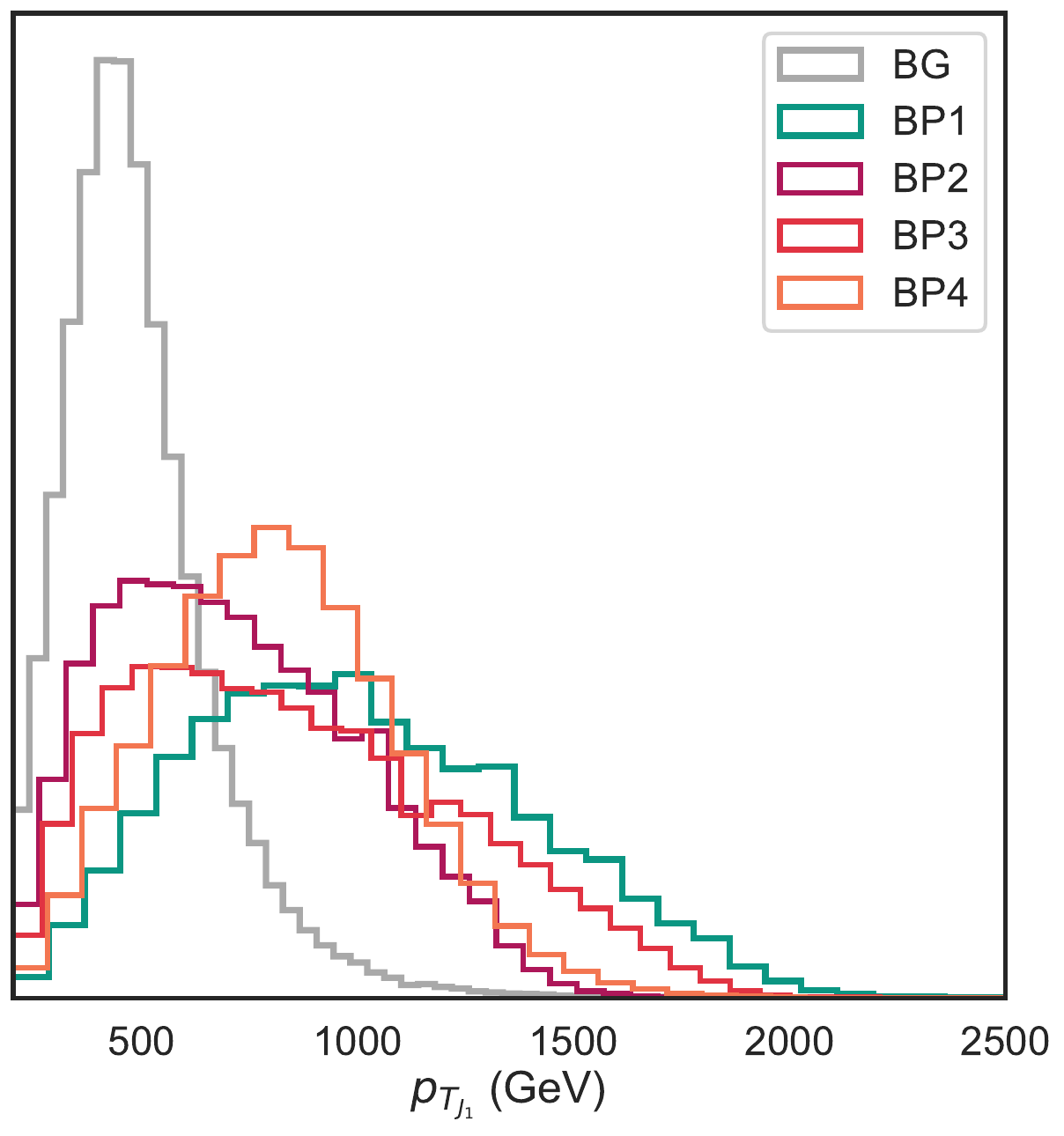}\label{fig:featureC}}\\
\subfloat[(d)]{\includegraphics[width=0.32\textwidth]{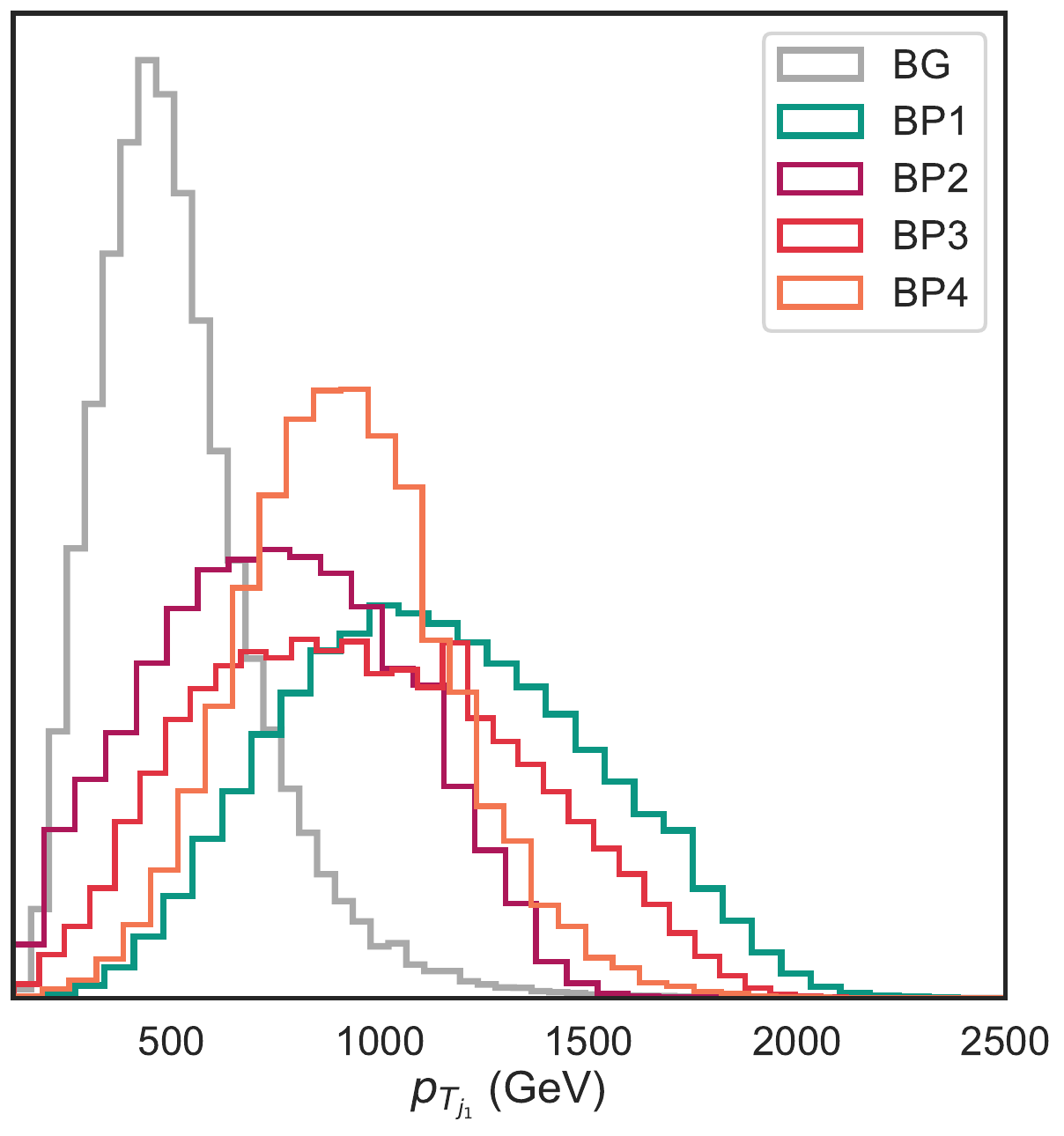}\label{fig:featureD}}\quad
\subfloat[(e)]{\includegraphics[width=0.32\textwidth]{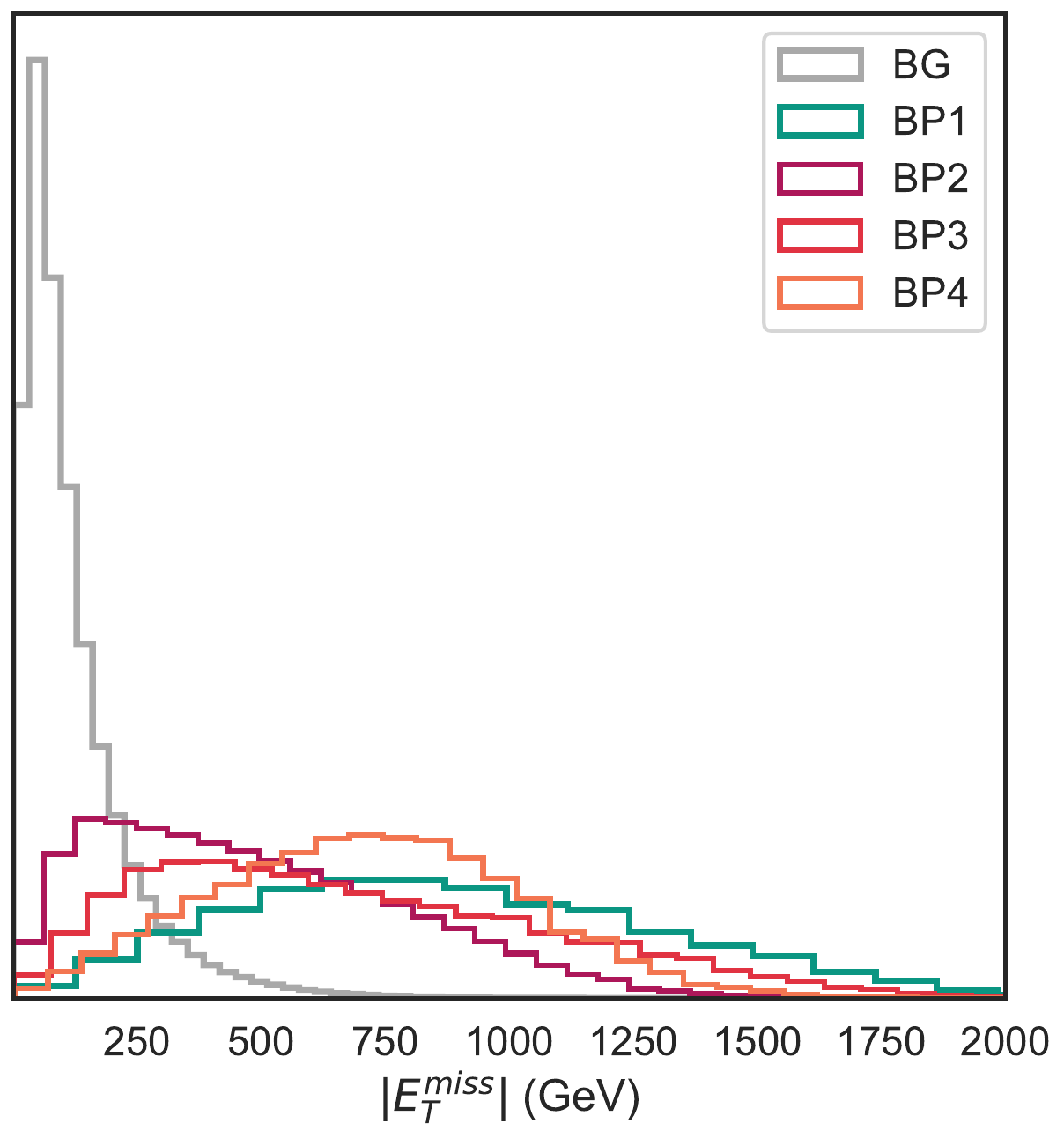}\label{fig:featureG}}\quad
\subfloat[(f)]{\includegraphics[width=0.305\textwidth]{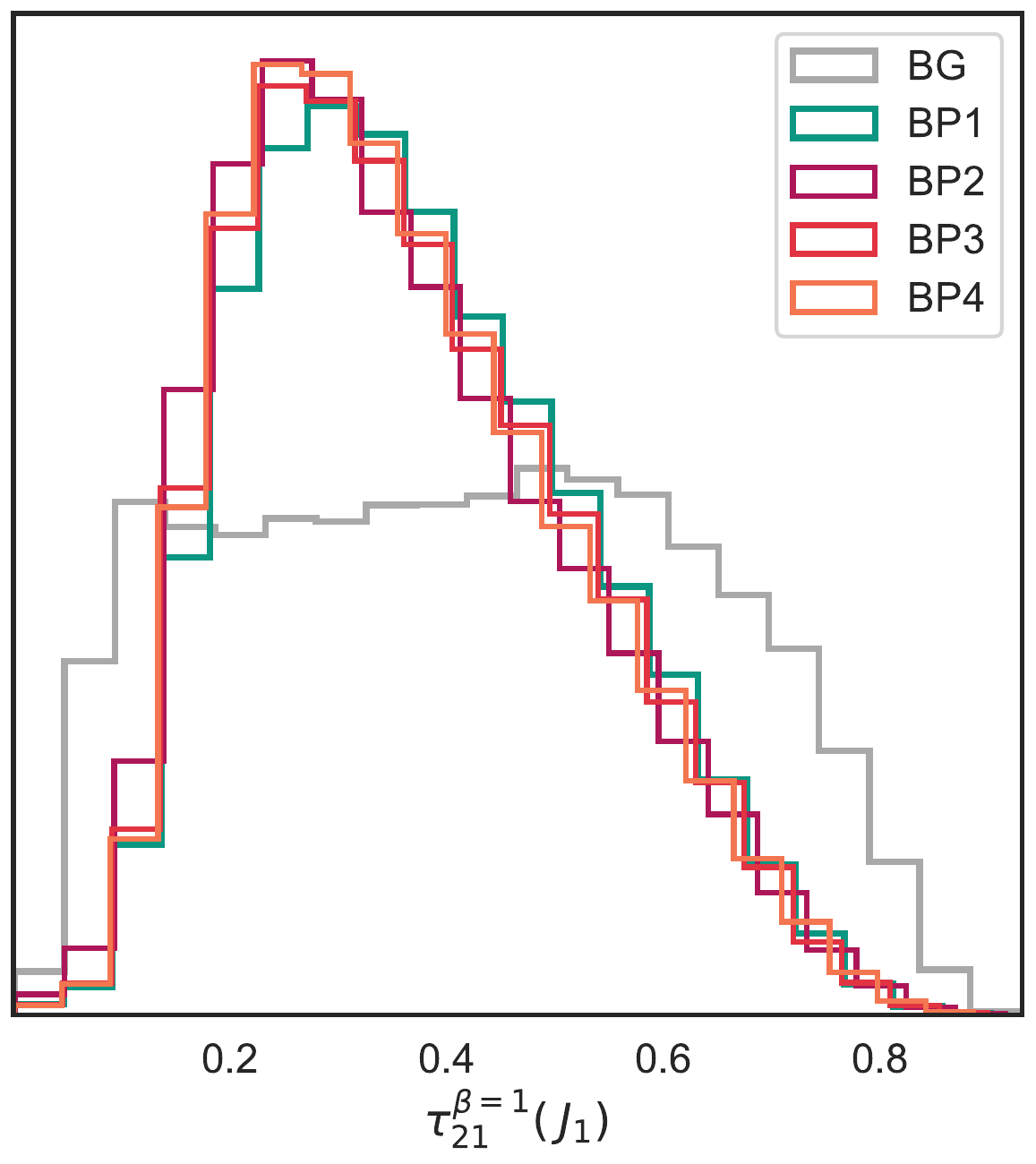}\label{fig:featureI}}\\
\subfloat[(g)]{\includegraphics[width=0.32\textwidth]{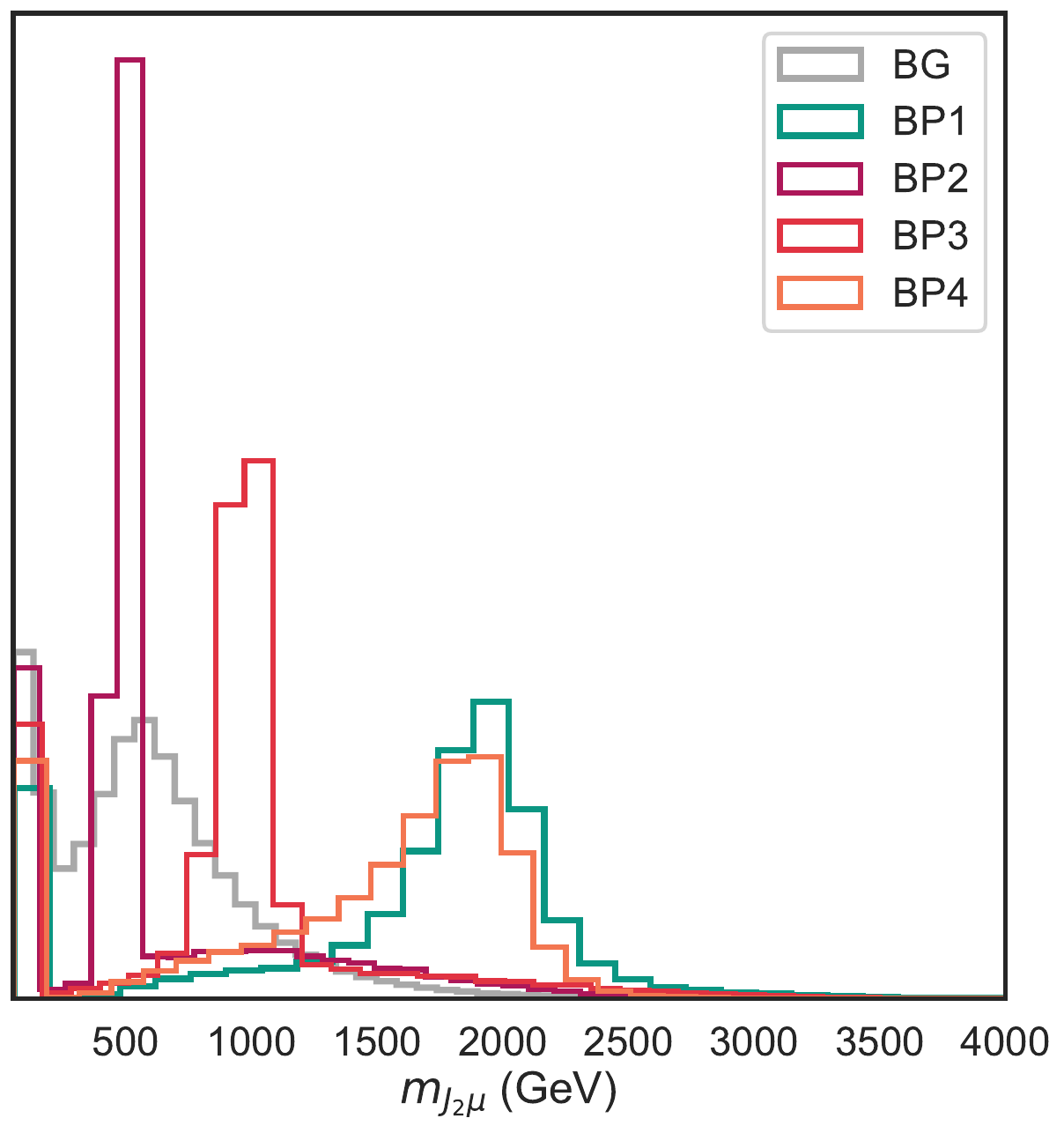}\label{fig:featureF}}\quad
\subfloat[(h)]{\includegraphics[width=0.315\textwidth]{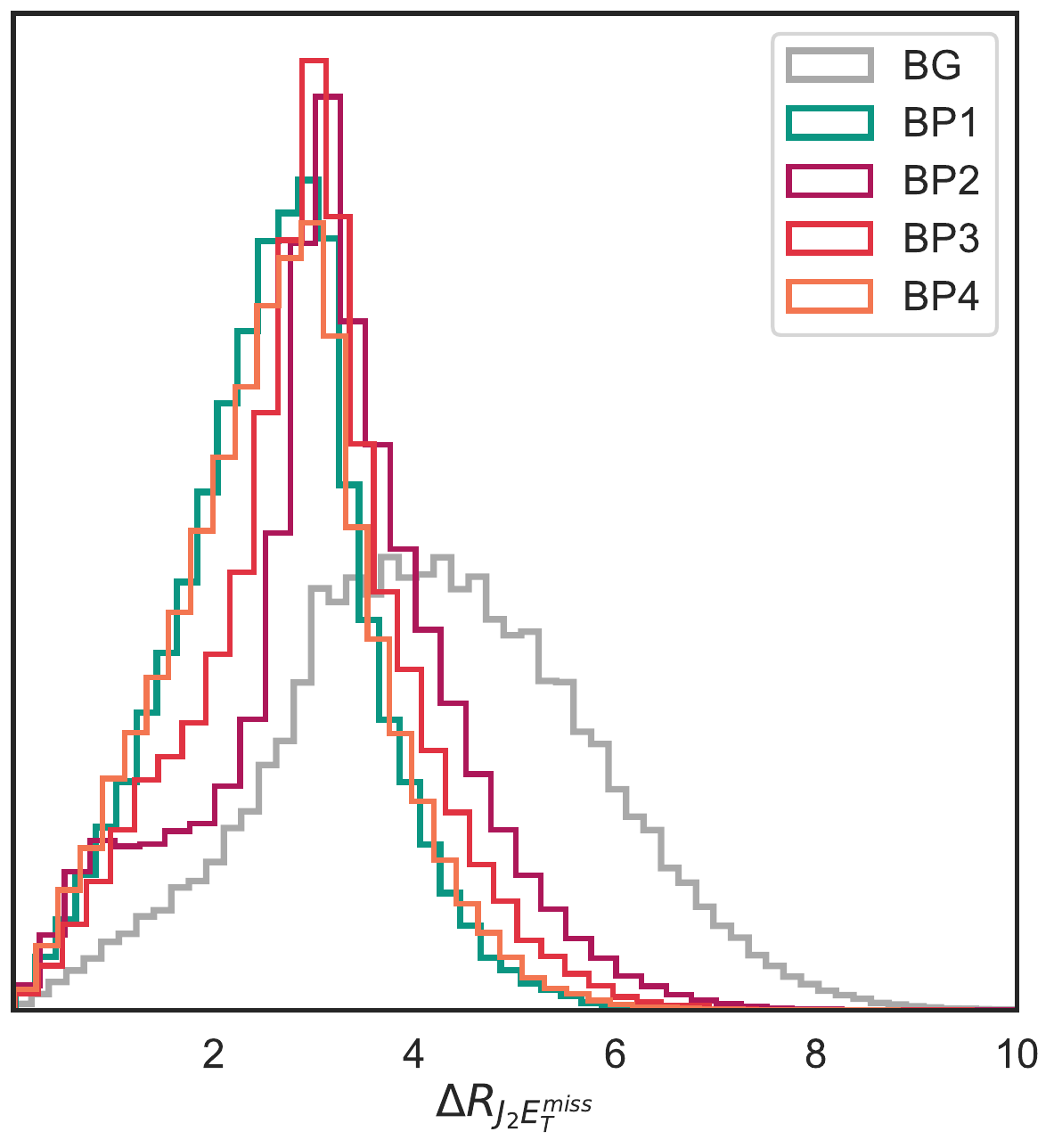}\label{fig:featureH}}\quad
\subfloat[(i)]{\includegraphics[width=0.315\textwidth]{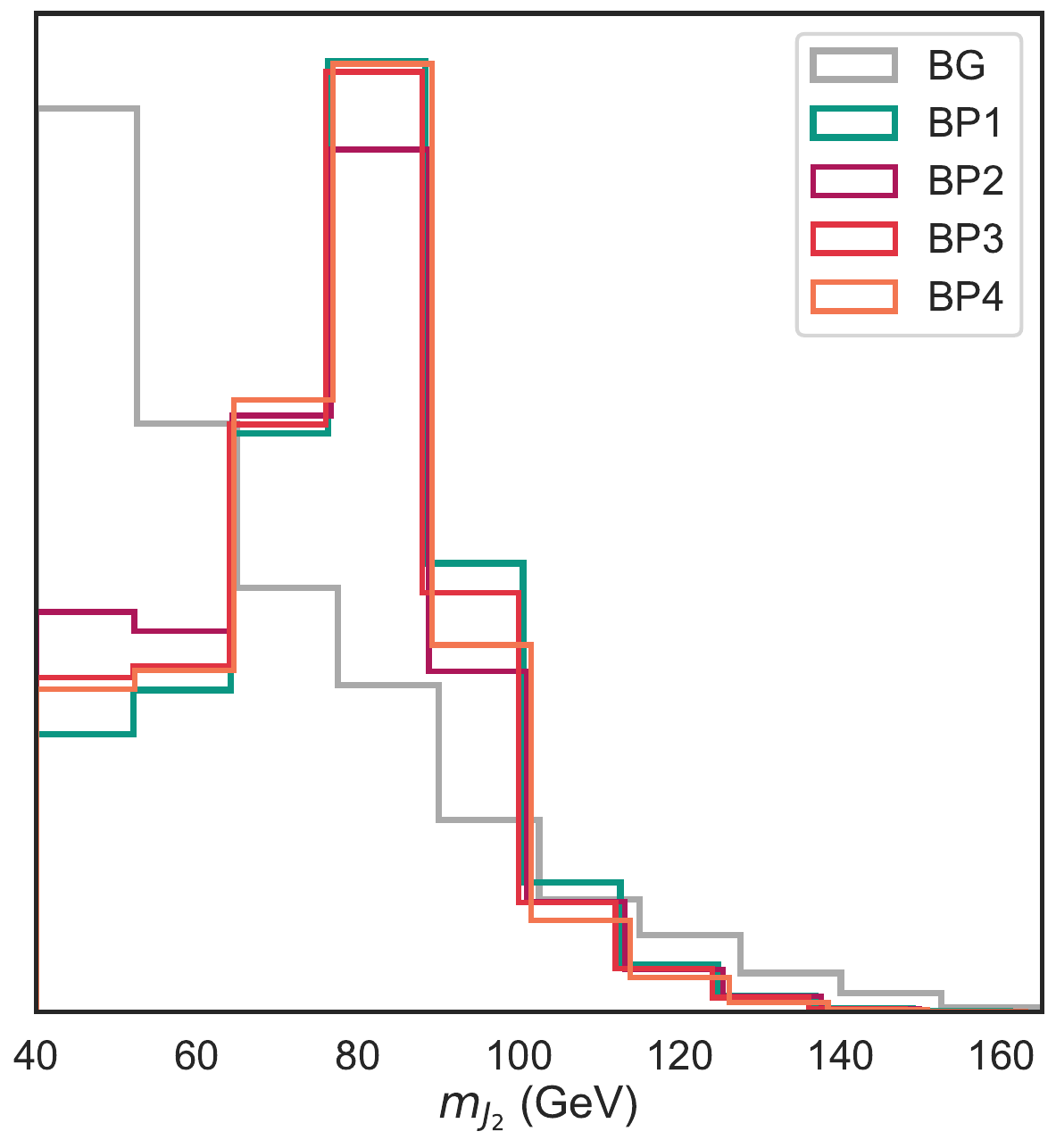}\label{fig:featureE}}\\
\caption{Normalised density plots of the key features after passing through the cuts $\mathfrak C_1$ to $\mathfrak C_5$ for four benchmarks points: (a) BP1 $\equiv (M_{Z'}=5.0,~M_{N_R}=2.0)$ TeV, (b) BP2 $\equiv (M_{Z'}=3.0,~M_{N_R}=0.5)$ TeV, (c) BP3 $\equiv (M_{Z'}=4.0,~M_{N_R}=1.0)$ TeV, and (d) BP4 $\equiv (M_{Z'}=4.0,~M_{N_R}\approx2.0)$ TeV. BP1 distributions are highlighted in green as we use this benchmark to tune the DNN model. Here, the fatjets are mass-ordered and the jets are $p_T$-ordered. }\label{fig:inputvars_monoLep}
\end{figure*}

\begin{table}[t]
\caption{Effective $U(1)_z$ charges for various $E_6$ embeddings and the GS model. For the six $E_6$ embeddings, we compute the charges using the leptophobic $\tan\theta$ and $\delta$ combinations from Table~\ref{tab:chargeGUT}. For the GS model, we use the benchmark point mentioned in Sec.~\ref{subsec:LPGS}.\label{tab:z-charges}}
\centering{\renewcommand\baselinestretch{1.5}\selectfont
\begin{tabular*}{\columnwidth}{ c@{\extracolsep{\fill}} rrr}
\hline
Model&$z_u$ & $z_d$ & $z_N$ \\
\hline\hline
$E_6$ embedding\\\hline
I & $1.242$ & $0.786$ & $-1.667$ \\
II & $1.242$ & $0.786$ & $-1.667$ \\
III & $0.393$ & $0.248$ & $-0.527$ \\
IV & $0.962$ & $0.481$ & $-1.936$ \\
V & $0.481$ & $0.481$ & $-1.936$ \\
VI & $0.393$ & $0.248$ & $-0.527$ \\
\hline
GS &$0.707$ &$0.707$ &$6$\\
\hline
\end{tabular*}}
\end{table}

\begin{table*}
\caption{Projected number of events at the $14$ TeV HL-LHC at different stages of analysis in the monolepton channel for the two benchmark points, BP1 $ \equiv (M_{Z'} = 5.0,~M_{N_R} = 2.0)$ TeV and BP2 $ \equiv (M_{Z'} = 3.0,~M_{N_R} = 0.5)$ TeV (see Fig.~\ref{fig:inputvars_monoLep}) with $g_z = 0.3$. Here,  $S_1$ and $S_2$ denote the number of signal events for BP1 and BP2, respectively. We obtain DNN(BP1) and DNN(BP2) after training the network on BP1 and BP2 training data, respectively. For the signal, we assume a $K$ factor of $1.3$~\cite{Eichten:1984eu} and BR$(Z'\to N_R N_R)=50\%$.\label{tab:monolepCutFlow}}
\centering{\renewcommand\baselinestretch{1.5}\selectfont
\begin{tabular*}{\textwidth}{ c@{\extracolsep{\fill}} r r r r r r r r r}
 \hline
 & \multicolumn{2}{c}{Signal}&\multicolumn{5}{c}{Background}\\\cline{2-3}\cline{4-8}
  & $S_1$ & $S_2$ & $t_h t_\ell$ & $t_\ell t_\ell$ & $W_\ell + j$ & $W_\ell W_h + j$ & $Z_h W_\ell + j$ & $S_1/B$ & $S_2/B$\\ [0.5ex] 
 \hline \hline
 Initial & $235$ & $12409$ & $276832$ & $4382$ & $392206$ & $291684$ & $22729$ & $2.37 \times 10^{-4}$ & 
 $1.26 \times 10^{-2}$ \\ 
 \hline
 $\mathfrak C_1$ & $234$ & $11764$ & $262336$ & $4035$ & $379874$ & $284631$ & $22168$ & $2.46 \times 10^{-4}$ & $1.23 \times 10^{-2}$  \\ 
 $\mathfrak C_2$ & $186$ & $9431$ & $165438$ & $1006$ & $271623$ & $209486$ & $16168$ & $2.80 \times 10^{-4}$ & $1.42 \times 10^{-2}$ \\
 $\mathfrak C_3$ & $184$ & $9158$ & $164646$ & $964$  & $257573$ & $205999$ & $15973$ & $2.85 \times 10^{-4}$ & $1.41 \times 10^{-2}$ \\
 $\mathfrak C_4$ & $152$ & $5224$ & $101562$ & $711$  & $105221$ & $99886$  & $7559$  & $4.83 \times 10^{-4}$ & $1.65 \times 10^{-2}$ \\
 $\mathfrak C_5$ & $151$ & $5212$ & $96071$  & $683$  & $98984$  & $94865$  & $7171$  & $5.07 \times 10^{-3}$ & $1.66 \times 10^{-2}$ \\
 \hline
 DNN(BP1) & $147$ & --- & $\sim 11$ & $\sim 1$ & $\sim 136$ & $\sim 213$  & $\sim 15$ & $0.39$ & --- \\ 
 DNN(BP2) &  ---  & $4534$ & $\sim 8$ & $\sim 1$ & $\sim 82$ & $\sim 160$  & $\sim 10$ & --- & $17.37$ \\ [1ex] 
 \hline
\end{tabular*}}
\end{table*}
\begin{table}[!t]
\caption{Kinematic variables chosen to analyse the monolepton signature; $\tau$'s are the n-subjettiness ratios.\label{tab:monolep_varList}}
\centering{\renewcommand\baselinestretch{1.5}\selectfont
\begin{tabular*}{\columnwidth}{c @{\extracolsep{\fill}} l} 
\hline
Variable & Description \\
\hline \hline
$H_T$ & Sum of transverse momenta of hadronic objects.\\
$p_{T_i}$ & Transverse Momenta of all reconstructed objects.\\
$\slashed{E}_T$ & Missing transverse energy. \\
$m_{J_i}$ & Invariant Masses of the reconstructed fatjets. \\
$\Delta R_{ik}$ & Distances between any two reconstructed objects.\\
$m_{J_i \mu}$ & Invariant masses of the fatjet-lepton pairs.\\ 
$m_{j_i j_k}$ & Invariant masses of any two reconstructed AK-4 jets.\\ 
$\tau_{21}^{\beta=1}$ & $\tau_2/\tau_1$, $\beta=1.0$ for each fatjet. \\
$\tau_{21}^{\beta=2}$ & $\tau_2/\tau_1$, $\beta=2.0$ for each fatjet.\\ 
$\tau_{32}^{\beta=1}$ & $\tau_3/\tau_2$, $\beta=1.0$ for each fatjet.\\ 
$\tau_{32}^{\beta=2}$ & $\tau_3/\tau_2$, $\beta=2.0$ for each fatjet.\\ 
\hline
\end{tabular*}}
\end{table}

\section{Collider Analysis}\label{sec:collider}
\noindent
For collider analysis, we use the following simplified Lagrangian,
\begin{align}\label{effective_charge}
    \mc{L} \supset \dfrac{g_z}{2}\lt(z_u \overline{u}_i\gm^\mu u_i + z_d \overline{d}_i\gm^\mu d_i + z_N \overline{N}_R\gm^\mu N_R\rt)Z^\prime_\mu\ ,
\end{align}
where $z_{u,d}$ are the effective $U(1)_z$ charges of up and down type quarks (for any quark $q$, $z_{q}^2=z_{q_L}^2 + z_{q_R}^2$). We show $z_u$ and $z_d$ for the six different $E_6$ embeddings and the GS model in Table~\ref{tab:z-charges}. Between the two colour-singlet isosinglets in Table~\ref{tab:chargeGUT}, we identify the one with higher absolute $U(1)_z$ charge with the RHN, i.e., $z_N=~\mbox{Max}(z_\n,z_{\widetilde\n})$.

We use \textsc{FeynRules}~\cite{Alloul:2013bka} to build the above model and obtain the Universal FeynRules Output (UFO). We simulate the hard scattering in \textsc{MadGraph5}~\cite{Alwall:2014hca}, showering and hadronisation  in \textsc{Pythia}~\cite{Sjostrand:2014zea} and the LHC detector environment in  \textsc{Delphes3}~\cite{deFavereau:2013fsa}. The events are generated at $\sqrt{s} = 14$ TeV.  Jets are clustered using the anti-$k_T$ algorithm~\cite{Cacciari:2008gp} with $R=0.4$ (we call them AK-4 jet). For the electron, the \texttt{DeltaRMax} parameter in the \textsc{Delphes} card (distance between the other isolated objects and the identified electron)  for isolation has been modified to $0.2$ from $0.5$.

\subsection{Event selection criteria: Monolepton channel}\label{sec:monolep-evsel}
\noindent We show the signal topology in  Fig.~\ref{fig:ZprimeFeynDiag}. In Ref.~\cite{Arun:2022ecj}, we performed a cut-based analysis of the dilepton signature. Here, in this paper, we mainly focus on the  monolepton signature, which arises when one of the RHNs decays to a hadronically decaying $W$ boson and a lepton (muon) while the other decays to a $Z/H$ boson and a neutrino:
\begin{equation*}
    p p \to Z' \to N_R N_R \to \m + \slashed{E} + \text{jets}.
\end{equation*}
 As explained earlier, probing the monolepton channel is more challenging than the dilepton case as the backgrounds (see Table~\ref{tab:monolepCutFlow}) are harder to tame with simple kinematic cuts. Hence, we employ a DNN model to enhance the signal sensitivity. We look at $Z'$ masses between $3-7$ TeV and RHNs ranging from $0.5$ to $3.5$ TeV. From the topology and the distributions of the final state, we design the following basic selection criteria for the signal events:
\begin{itemize}
    \item[$\mathfrak C_1$:] $H_T > 750$~GeV, where $H_T$ is the scalar sum of the transverse momenta of all hadronic objects in an event.
    \item[$\mathfrak C_2$:] Exactly one muon, with $p_T > 120$ GeV with $|\eta| < 2.5$. (We consider the muon as it has the best detector sensitivity among the leptons.)
    
    \item[$\mathfrak C_3$:] At least two AK-4 jets. We also demand that the leading jet has $p_T >  120$ GeV and the subleading one has $p_T > 40$ GeV. 
    
    \item[$\mathfrak C_4$:] Exactly two fatjets, clustered using the anti-$k_T$ algorithm with $R=0.7$ and $p_T>200$ GeV with mass $M_J \in [40,165]$.

    \item[$\mathfrak C_5$:] The invariant mass between the reconstructed muon and missing transverse energy ($\slashed E_T$) must be greater than $140$ GeV, i.e., $M_{\mu \slashed E_T} > 140$ GeV.
\end{itemize}
We pick the fatjet radius $R=0.7$ after optimising to tag a boosted two-pronged jet. The constraints on fatjet mass ($\mathfrak{C}_4$) are chosen to include reconstructed $W,\ Z$ and $H$ fatjets. Since the missing energy and muon come from the decay of $W$-boson in all the backgrounds, we put a high cut on the invariant mass of reconstructed lepton and missing transverse energy pair in $\mathfrak{C}_5$.

\begin{figure*}
\captionsetup[subfigure]{labelformat=empty}
\centering
\subfloat[(a)]{\includegraphics[width=0.4\textwidth]{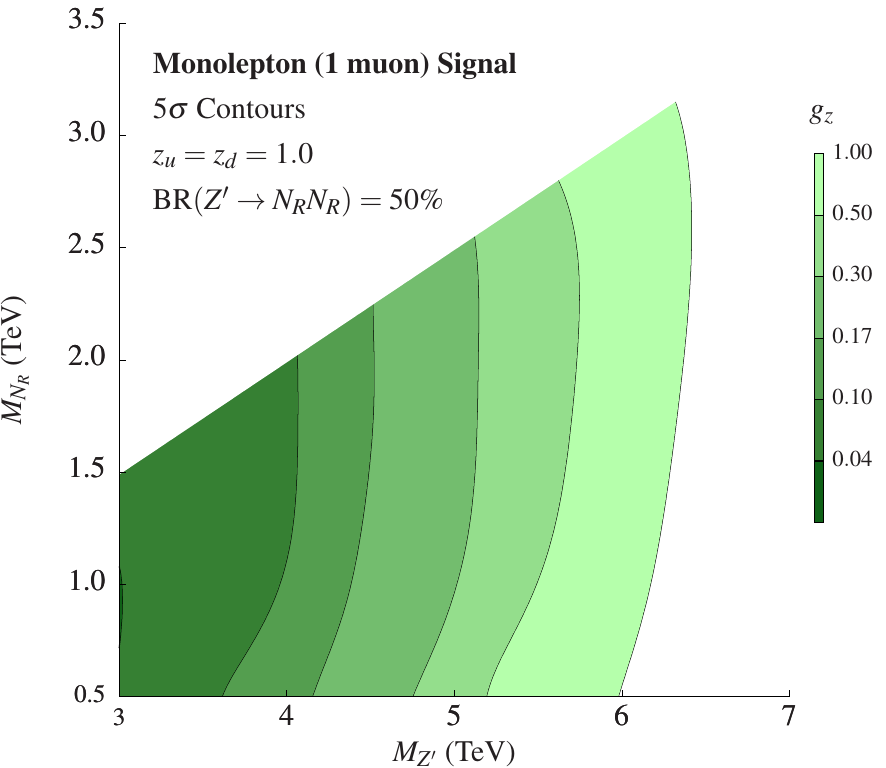}}\hspace{1cm}
\subfloat[(b)]{\includegraphics[width=0.4\textwidth]{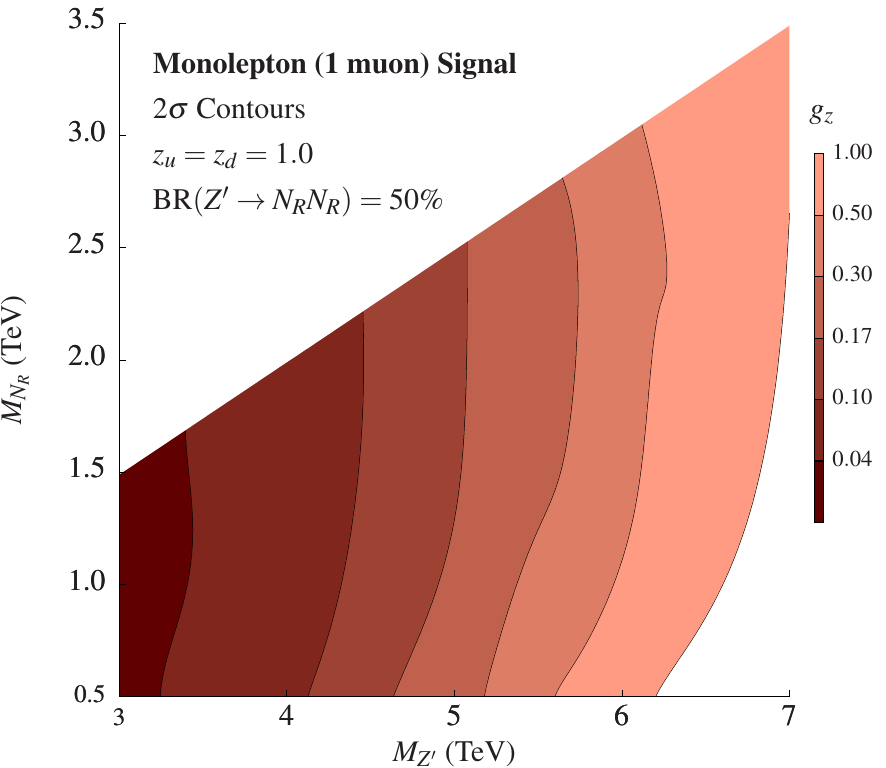}}
\caption{The (a) discovery ($5 \sigma$) and (b) exclusion ($\sim 2\sigma$) contours for the monolepton (1-muon) channel of $pp\to Z' \to N_R N_R$ at HL-LHC.}
\label{fig:ZcontoursM}
\end{figure*}
\begin{figure*}
\centering
\captionsetup[subfigure]{labelformat=empty}
\subfloat{\includegraphics[width=0.27\textwidth]{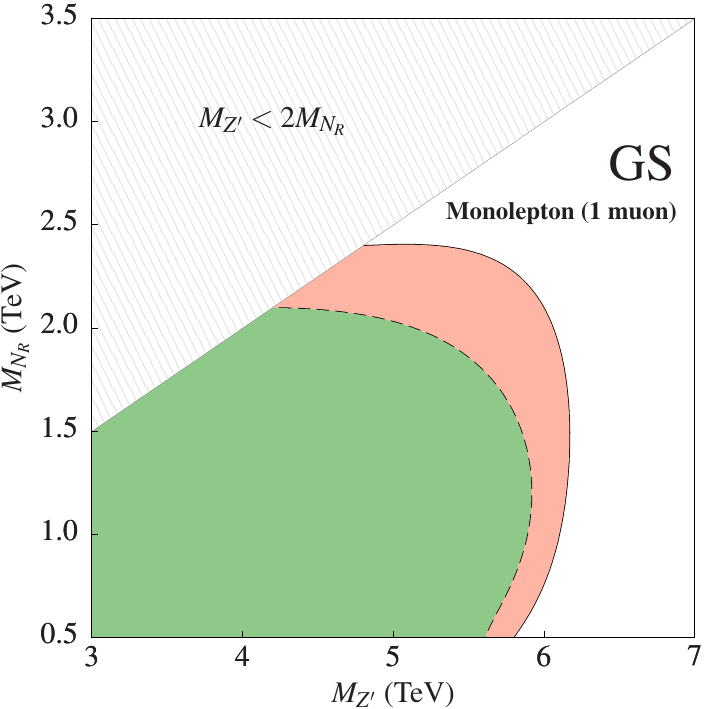}}\label{fig:ml_gs}\hspace{1cm}
\subfloat{\includegraphics[width=0.27\textwidth]{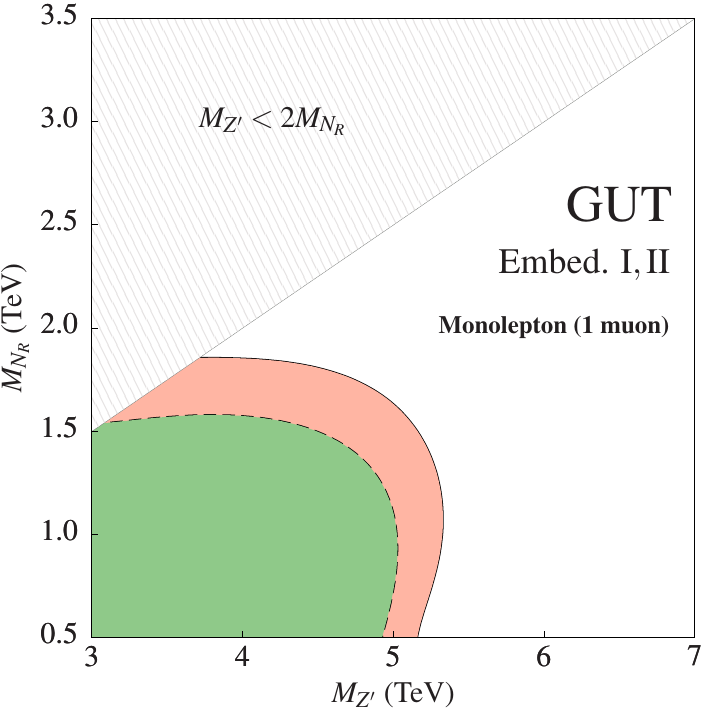}\label{fig:ml_gut12}}\hspace{1cm}
\subfloat{\includegraphics[width=0.27\textwidth]{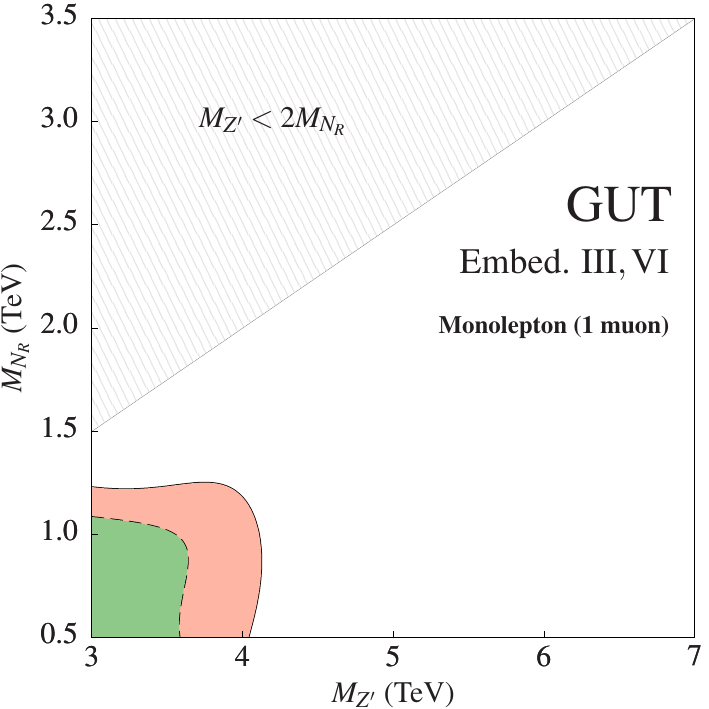}\label{fig:ml_gut36}}\\
\subfloat{\includegraphics[width=0.27\textwidth]{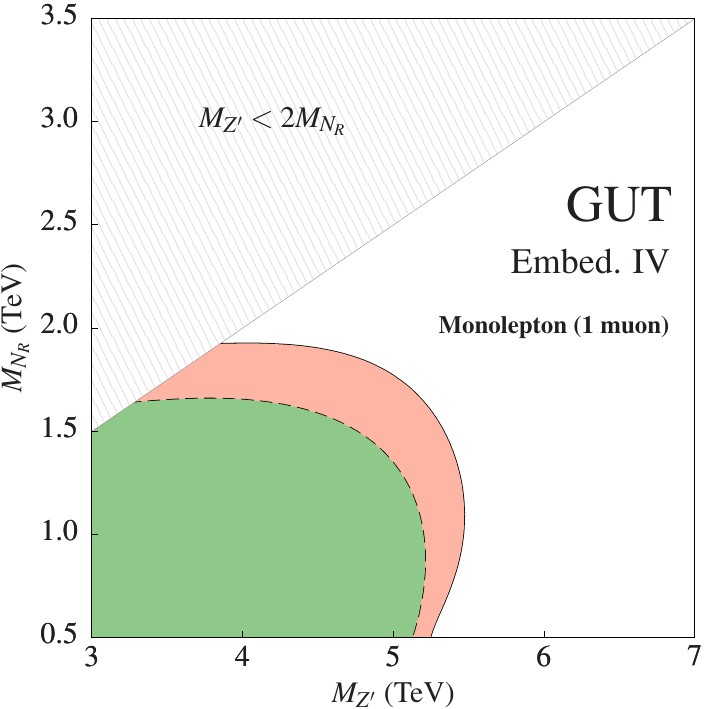}\label{fig:ml_gut4}}\hspace{1cm}
\subfloat{\includegraphics[width=0.27\textwidth]{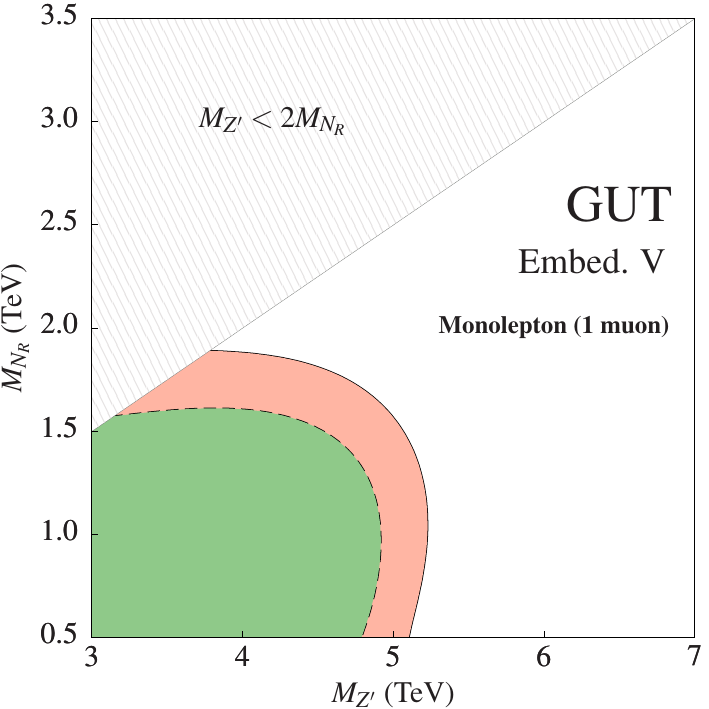}\label{fig:ml_gut5}}
\caption{HL-LHC reaches for the GS model and the $E_6$ embeddings in the monolepton channel for $g_z=0.72$. The green regions (dotted contours) are discoverable ($5\sigma$), and the red regions are excludable ($2\sigma$). The $U(1)_z$ charges are found in Table~\ref{tab:z-charges} .}\label{fig:GUTReachM}

\captionsetup[subfigure]{labelformat=empty}
\subfloat[]{\includegraphics[width=0.27\textwidth]{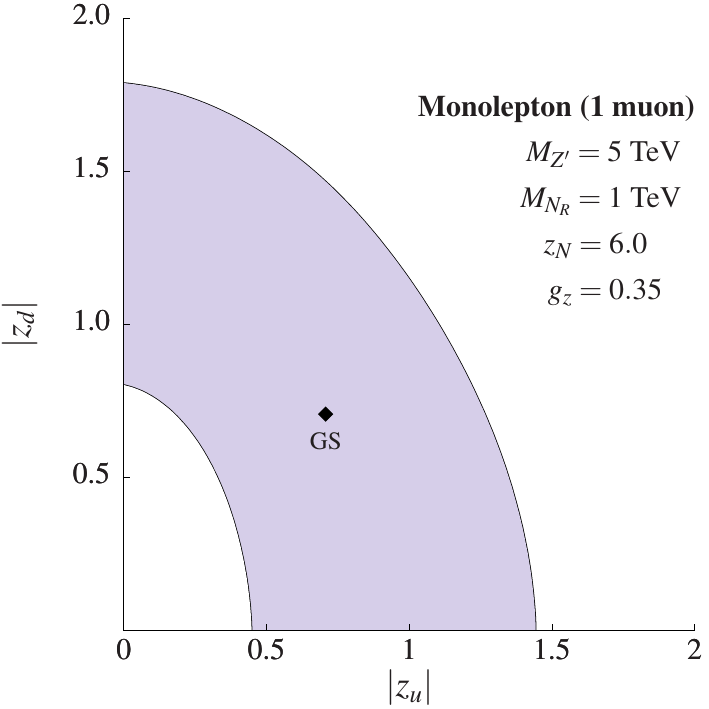}\label{fig:RHNreach_gs}}\hspace{1cm}
\subfloat[]{\includegraphics[width=0.27\textwidth]{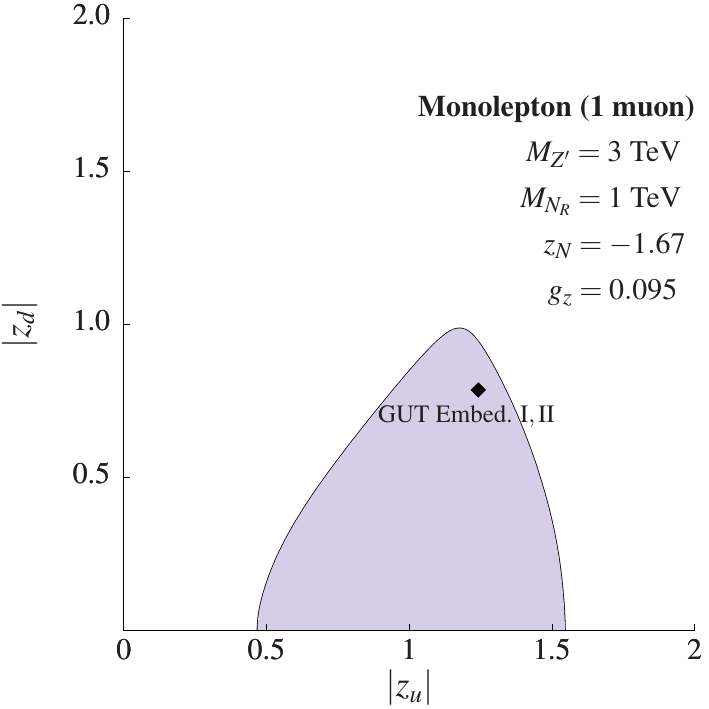}\label{fig:RHNreach_gut12}}\\
\subfloat[]{\includegraphics[width=0.27\textwidth]{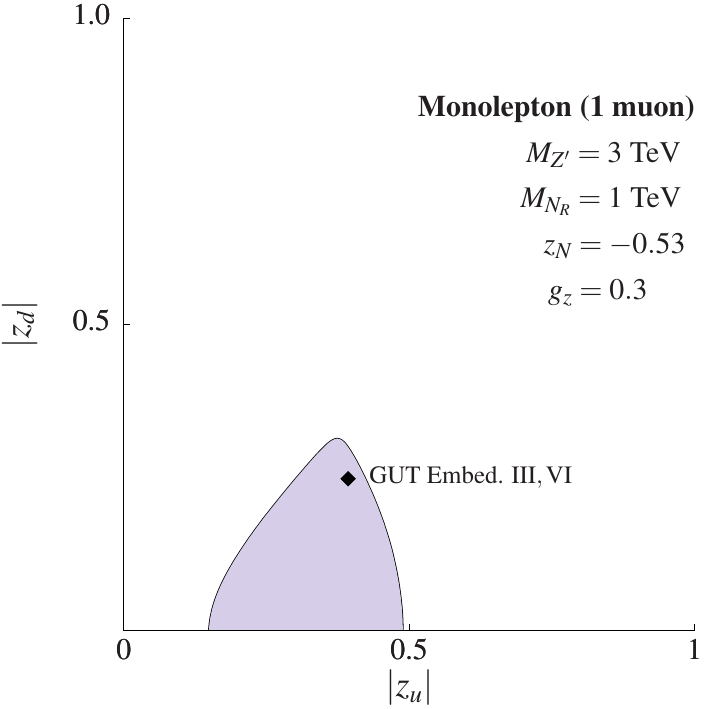}\label{fig:RHNreach_gut36}}\hspace{1cm}
\subfloat[]{\includegraphics[width=0.27\textwidth]{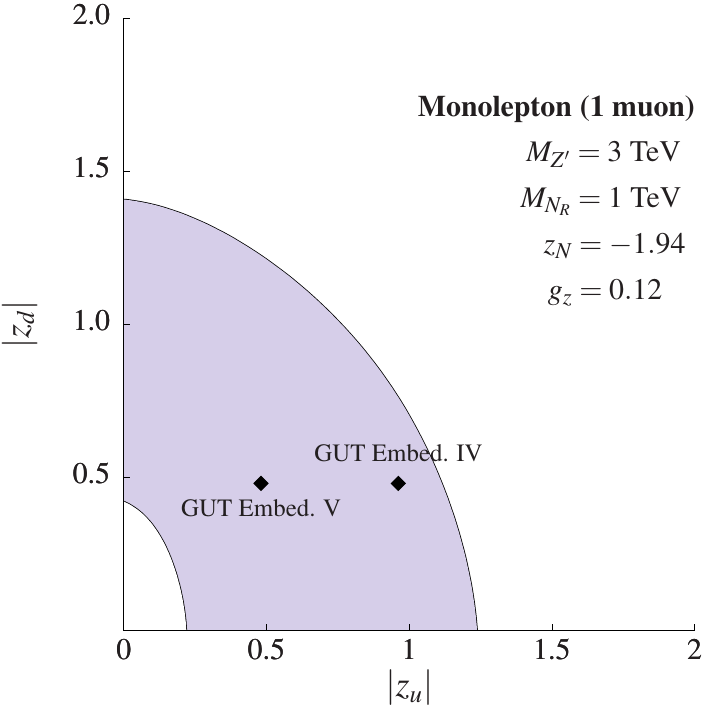}\textbf{}\label{fig:RHNreach_gut4}}
\caption{Regions where the monolepton signal achieves at least $2\sigma$ significance at the HL-LHC but are beyond the projected reach in the dijet channel (see text). \label{fig:GUT-GS-RHNreach}}
\end{figure*}

\begin{figure*}[]
\captionsetup[subfigure]{labelformat=empty}
\centering
\subfloat[(a)]{\includegraphics[width=0.4\textwidth]{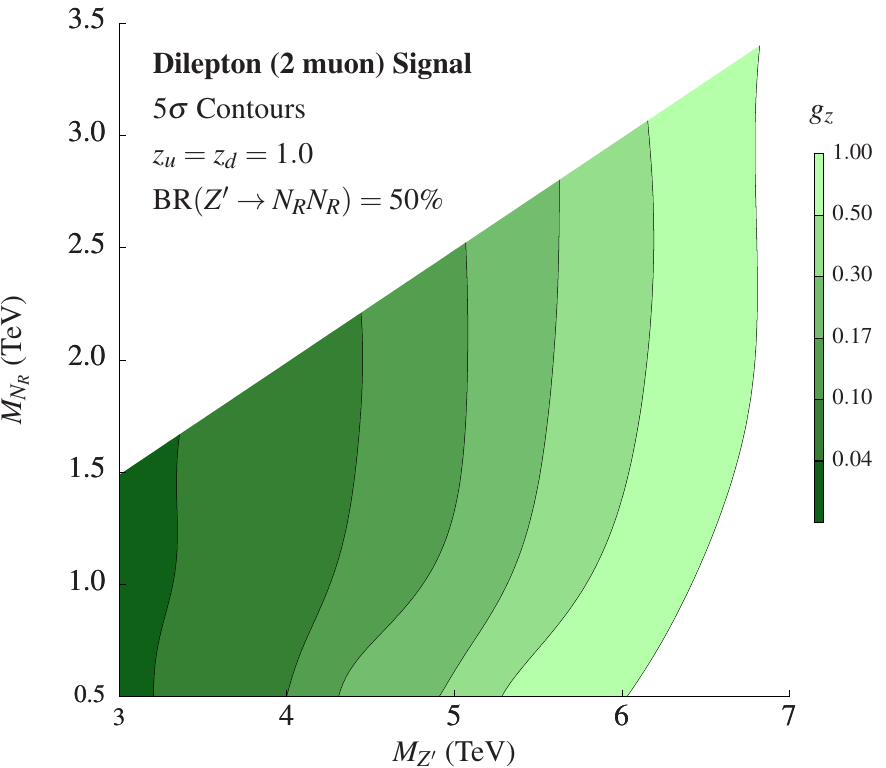}}\hspace{1cm}
\subfloat[(b)]{\includegraphics[width=0.4\textwidth]{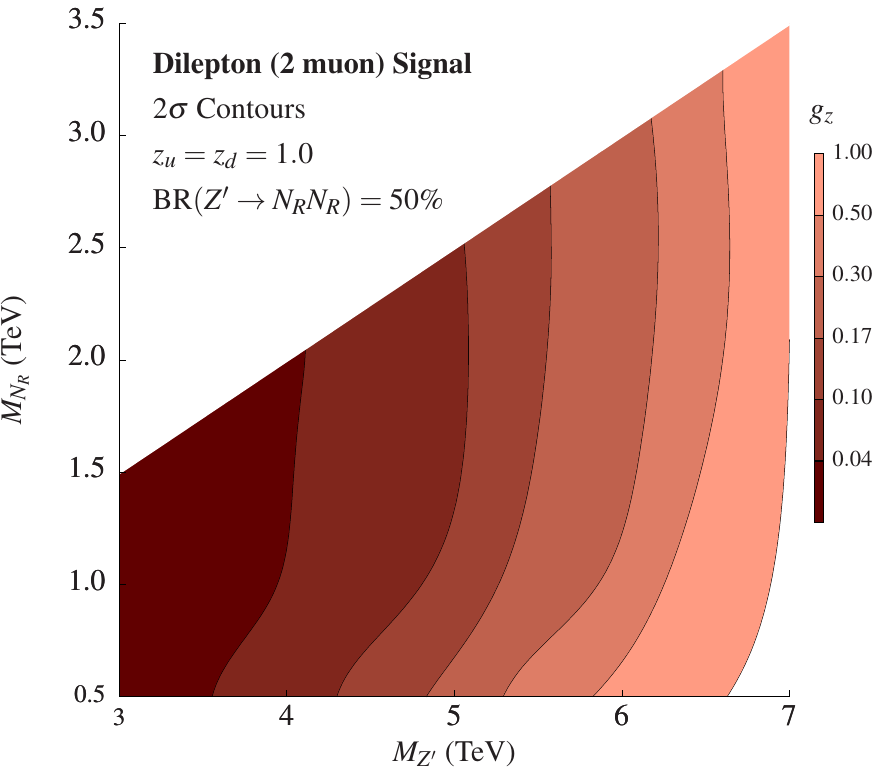}}
\caption{The (a) discovery ($5 \sigma$) and (b) exclusion ($\sim 2\sigma$) contours for the dilepton (2-muon) channel of $pp\to Z' \to N_R N_R$ at HL-LHC.}
\label{fig:ZcontoursD}
\captionsetup[subfigure]{labelformat=empty}
\subfloat{\includegraphics[width=0.27\textwidth]{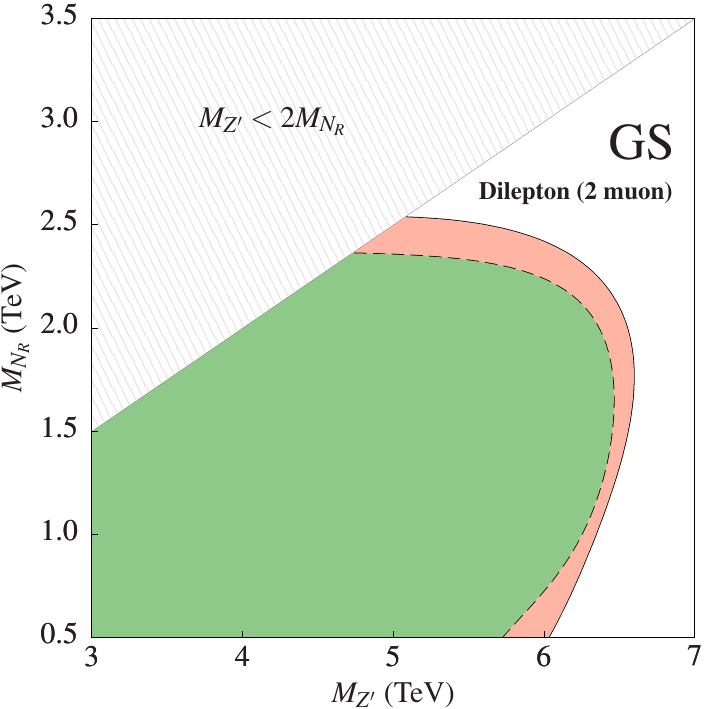}}\label{fig:dl_gs}\hspace{1cm}
\subfloat{\includegraphics[width=0.27\textwidth]{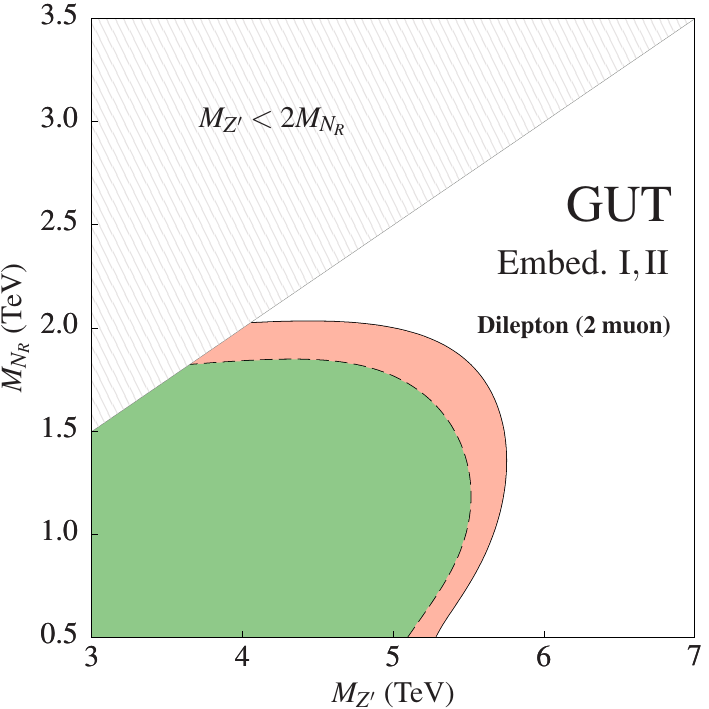}\label{fig:dl_gut12}}\hspace{1cm}
\subfloat{\includegraphics[width=0.27\textwidth]{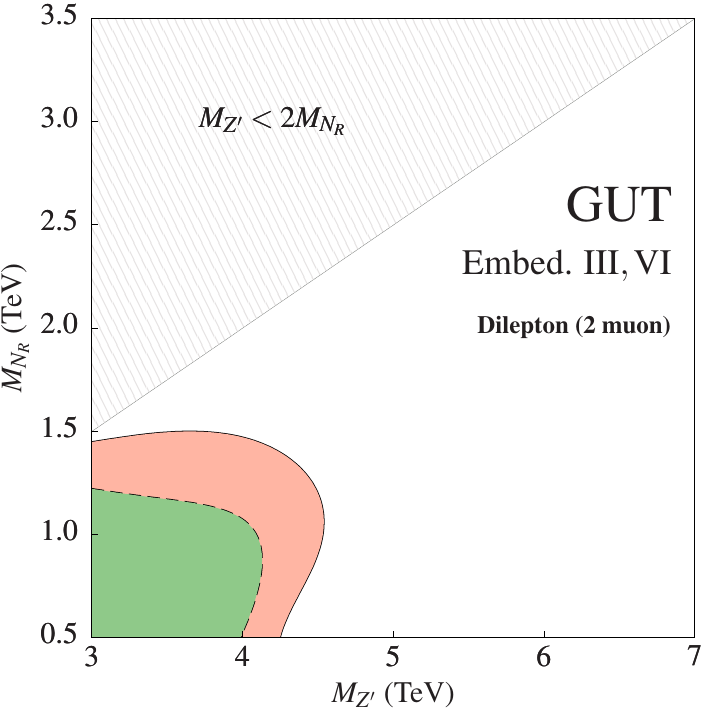}\label{fig:dl_gut36}}\\
\subfloat{\includegraphics[width=0.27\textwidth]{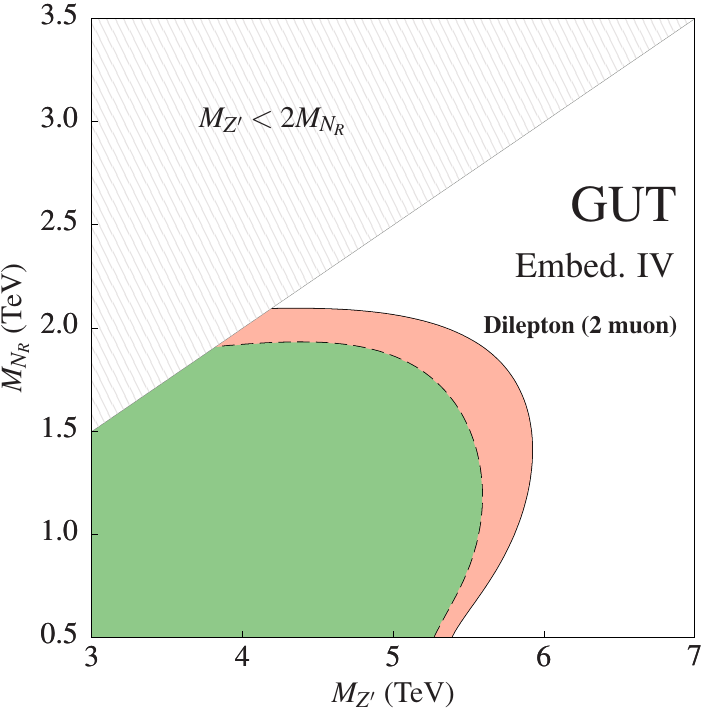}\label{fig:dl_gut4}}\hspace{1cm}
\subfloat{\includegraphics[width=0.27\textwidth]{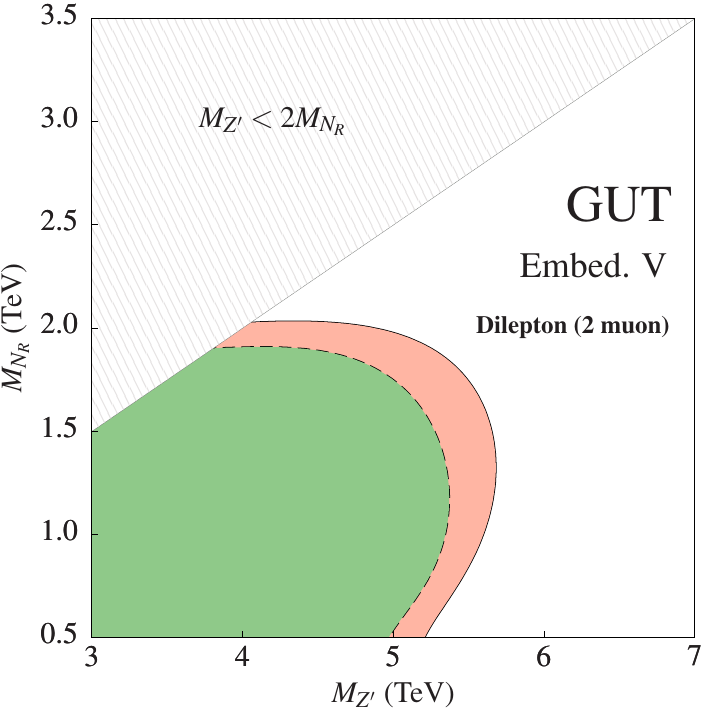}\label{fig:ml_gut5}}\\
\caption{HL-LHC reaches for the GS model and the $E_6$ embeddings in the dilepton channel for $g_z=0.72$. The green regions (dotted contours) are discoverable ($5\sigma$), and the red regions are excludable ($2\sigma$). The $U(1)_z$ charges are found in Table~\ref{tab:z-charges} .}\label{fig:GUTReachD}
\end{figure*}

\subsection{Background processes}
\noindent The major backgrounds considered for the analysis are listed in Table~\ref{tab:monolepCutFlow}, along with the cut efficiencies and the total number of events at the HL-LHC luminosity, $3$ ab$^{-1}$. We generate the background events with some generation-level cuts to reduce the computation time. The \emph{$W^{\pm}$ + jets} process is the leading background before the cuts; the demand of a high-$p_T$ jet ($\mathfrak C_3$) and two fatjets ($\mathfrak C_4$) reduces this background significantly. The contributions from the \emph{$W^+ W^-$ + jets} process drop drastically with the demand of two fatjets ($\mathfrak C_4$). The fatjet criterion ($\mathfrak C_4$) also cuts the \emph{$t\bar{t}$ + jets} background as most of these events have just one fatjet. The hadronically-decaying top is sometimes reconstructed as the fatjet in a fraction of \emph{$t\bar{t}$ + jets} events; the demand on the fatjet mass cuts down these events. After the event selection criteria are enforced, \emph{$W^+ W^-$ + jets} and \emph{$t\bar{t}$ + jets} become the leading backgrounds.

\subsection{Kinematic picture and the Deep Learning model}
\noindent We use the kinematic information of the identified objects, i.e.,  one muon, three AK-4 jets, and two fatjets, along with some derived quantities to create the input set of variables for the multivariate analysis---the complete list of variables is given in Table~\ref{tab:monolep_varList} where $J$ is used to denote a fatjet whereas $j$ is used to identify an AK-4 jet. 

We show the distributions of the key features of the signal and the background for three benchmark parameters in Fig.~\ref{fig:inputvars_monoLep}. The $H_T$ (scalar sum of all hadronic objects) distributions show a clear separation between the signals and background due to the boosted nature of the final state; see Fig.~\ref{fig:featureA}. Moreover, for the signals,  the transverse momentum distributions of the tagged muon, AK-4 jets, and the fatjet and the $\slashed E_T$ distribution peak at considerably higher values than those of the background, see Figs.~\ref{fig:featureB},~\ref{fig:featureC},~\ref{fig:featureD},~\ref{fig:featureG}. We also consider the $n$-subjettiness variables~\cite{Thaler:2010tr} which check for two-pronged ($\tau_2/\tau_1$) and three-pronged ($\tau_3/\tau_2$) nature of a fatjet for various values of $\beta$ ($\beta \in \left\{ 1,2,3\right\}$). Since our fatjet parameters are tuned to identify a $W$-like (two-pronged) jet, the $\tau_{2}/\tau_{1}$ ratios are the key variables. Of these, the $\tau_{2}/\tau_{1}$ for $\beta=2$ gives a clear separation between the signals and the background [Fig.~\ref{fig:featureI}]. The $\tau_{3}/\tau_{2}$ ratios do not offer clear separations, but, nevertheless,  we include them due to the inclusive nature of the analysis cuts on the number of fatjets. The variable $m_{J_i \ell}$ (invariant mass of a fatjet and the tagged muon) reconstructs the mass of the RHN very well---the trend can be clearly seen in Fig.~\ref{fig:featureF}. The $\Delta R_{J_i \slashed{E}_T}$ distributions in Fig.~\ref{fig:featureH} can also distinguish the signals and the background well, as the final state objects come from the decay of an on-shell $Z'$. 

To perform the classification task, we use a fully connected DNN model, the details of which are described in Appendix~\ref{App:DNN}. We pick the benchmark point $(M_{Z'}, M_{N_R}) = (5.0, 2.0)$ TeV to optimise the hyperparameters using a grid search. The performance metric is the signal significance ($\mc Z$ score),
\begin{equation}
    \label{eq:Zscore}
    \mathcal{Z} = \sqrt{2\left(N_S + N_B\right)\ln\left(\frac{N_S+N_B}{N_B}\right) - 2N_S},
\end{equation}
where $N_S$ and $N_B$ are the numbers of signal and background events the network allows at $3$ ab$^{-1}$ luminosity.

\section{Monolepton Channel Prospects}\label{sec:monoprospects}
\noindent
Fig.~\ref{fig:ZcontoursM} shows the HL-LHC discovery ($5\sigma$) and exclusion ($\sim 2\sigma$) reaches through the monolepton channel in a model-independent manner. Assuming a fixed BR($Z^\prime\to N_R N_R$) $=50\%$,  we mark the contours of the $U(1)_z$ gauge coupling, $g_z$ [Eq.~\eqref{pqterm}], needed to achieve $5\sigma$ and $2\sigma$ significances on the $M_{Z'}-M_{N_R}$ plane. To calculate the number of events, $N_S$ and $N_B$, we have used the final efficiencies of the DNN model at the working point with DNN response $= 0.95$. The $g_z$ contours are broadly insensitive to the mass of the RHN. However, because of the high boosts, the event selection efficiency around $M_{N_R}=0.5$ TeV (or less) is lower than those at larger values. Hence, one needs a larger $g_z$ to achieve a high signal significance. This can be seen from the slightly deformed contours between $M_{N_R}=1.0$ TeV and $M_{N_R}=0.5$ TeV.

In particular models, however, the $Z^\prime$ BR will not be fixed. For example, in Fig.~\ref{fig:GUTReachM}, we show the $5\sigma$ and $2\sigma$ contours for the GS model and the six $E_6$ embeddings for $g_z=0.72$ ($\approx g_{ew}$); the corresponding $U(1)_z$ charges are found in Table~\ref{tab:z-charges}. As the  $M_{N_R}/M_{Z^\prime}$ ratio approaches the threshold value, $1/2$, BR($Z^\prime\to N_R N_R$) drops. This is why the contours in Fig.~\ref{fig:GUTReachM} turn inwards with increasing $M_{N_R}$ near the kinematic threshold (top-left edge).

In Fig.~\ref{fig:GUT-GS-RHNreach}, we show the regions that will be beyond the reach of the dijet channel at the HL-LHC (approximately) but can be reached ($\gtrsim 2\sigma$) by the monolepton signal. To estimate the projected dijet reach, we have simply scaled down the current ATLAS dijet exclusion limits by the square root of the luminosity ratio ($\sqrt{3000/140}\approx 4.65$), ignoring the difference between the centre-of-mass energies.

\section{Update: Dilepton Channel Prospects}\label{sec:diprospects}
\noindent 
We update our earlier cut-based estimates of the dilepton prospects~\cite{Arun:2022ecj} with the current DNN model. The dileptonic signal comes from the following process,
\begin{equation}
    pp \to Z'\to N_R N_R \to \left(W^+ \mu^-\right)\:\left(W^{-}\mu^- \right),
\end{equation} 
where the $W$ bosons decay hadronically. See Ref.~\cite{Arun:2022ecj} for a discussion on the important background processes. The event-selection criteria remain similar to those in Sec.~\ref{sec:monolep-evsel} with some minor changes: in $\mathfrak C_2$, we now demand two muons with $p_T > 120$ GeV and the new fatjet window in $\mathfrak C_4$ is now $\left[40,120\right]$ to only tag $W$-like fatjets. To suppress the huge Drell-Yan dilepton background, a Z-veto cut is necessary; we also change  $\mathfrak C_5$ to demand that the invariant mass of the muon pair $(m_{\ell \ell})$ be greater than 140 GeV.

We show the model-independent HL-LHC discovery ($5\sigma$) and exclusion ($\sim 2\sigma$) reaches through the dilepton channel in Fig.~\ref{fig:ZcontoursD} and the $5\sigma$ and $2\sigma$ contours for the GS model and the six $E_6$ embeddings with $g_z=0.72$ in Fig.~\ref{fig:GUTReachD}. Comparing Fig.~\ref{fig:ZcontoursD}  with Fig.~4 of Ref.~\cite{Arun:2022ecj}, we see the DNN has enhanced the reaches (discovery and exclusion) significantly.

\section{Conclusions and Discussions}\label{sec:conclu}
\noindent
In this paper, we studied the prospects of a correlated search of a heavy neutral gauge boson $Z'$ and the right-handed neutrino $N_R$ in leptophobic $U(1)$ extensions. In particular, we analysed an experimentally unexplored channel $pp\to Z'\to N_R N_R$ where the $N_R$ pair decays to a monomuon final state. While in Ref.~\cite{Arun:2022ecj} we only considered the GS model and the standard $E_6$ embedding, here, we study all possible embeddings of $E_6$ GUT models where a leptophobic $Z^\prime$ can dominantly decay to TeV-scale RHNs in large parts of the parameter space. Moreover, unlike the dilepton signature considered in our previous paper~\cite{Arun:2022ecj}, probing the monolepton signature is highly challenging due to the huge background from the SM processes, which are hard to reduce with a cut-based analysis. Hence, in this paper, we used a DNN model to isolate the signal from the SM background, providing orders of magnitude gains in significance scores. With this, we found both mono and dilepton channels to be comparable and highly promising; for an order-electroweak coupling $g_z$, the leptophobic $Z'$ with mass up to $6-7$ TeV can be discovered at the $14$ TeV HL-LHC. Therefore, this follow-up paper improves upon our previous study on several counts: it covers more model possibilities with a more sophisticated analysis. It also covers the complimentary signatures---mono and dilepton channels---to present a comprehensive view of the prospects of this exciting but experimentally unexplored process.

Our findings point to a few directions for follow-up investigations. Studying the phenomenology of the Yukawa, scalar, and neutrino sectors of the GS and GUT models we consider will be interesting. It will also be interesting to explore the parameter space where the RHNs can produce displaced vertices (as we pointed out in Ref.~\cite{Arun:2022ecj}) or form $N_R$-jets~\cite{Mitra:2016kov}. On the other hand, our analysis is generic as it mainly relies on the kinematic features of the mono and dilepton final states produced from RHN pairs. Hence, the results---presented in Figs.~\ref{fig:ZcontoursM} and~\ref{fig:ZcontoursD} in a mostly model-independent manner---can be mapped to other BSM scenarios not necessarily related to the leptophobic $Z^\prime$. For example, we could consider $(\overline{q}\,\Gm\, q)(\overline{N}_R\,\Gm\, N_R)$-type operators (where $\overline{\psi}\,\Gm\,\psi$ is a Dirac bilinear) in the $N_R$-effective field theory framework~\cite{delAguila:2008ir,Bhattacharya:2015vja,Liao:2016qyd,Alcaide:2019pnf,Mitra:2022nri} to produce RHN pairs at the LHC. Similarly, a pair of RHNs can also come from a leptoquark pair~\cite{Bhaskar:2023xkm}: $pp \rightarrow \ell_q \ell_q \rightarrow (N_R j)(N_R j)$. In such cases, where RHNs are resonantly produced in pairs, we can directly recast our results for comparable kinematics. For example, we can consider the simplest case of a charge-$1/3$ singlet scalar leptoquark. For $M_{\ell_q}=1.5$ TeV, $M_{N_R}=0.5$ TeV (the kinematics is roughly comparable to the benchmark point BP2, defined in Fig.~\ref{fig:inputvars_monoLep}) and BR($\ell_q\to N_Rj$) $=100\%$, both the monolepton and the dilepton channels give $\mc Z>5$ (i.e., they are discoverable). The prospects are better than those reported in Ref.~\cite{Bhaskar:2023xkm}, which used no machine learning techniques.

Finally, some studies on the leptophobic $Z'$ in supersymmetric models are available in the literature~\cite{Langacker:1998tc,Araz:2017wbp,Frank:2020kvp}. Generally, in such models, the $Z^\prime$ can decay to a pair of charginos, producing a dilepton plus missing energy signal at the LHC~\cite{Frank:2020kvp}: 
\begin{equation*}
p~p \to Z' \to \widetilde{\chi}^\pm\widetilde{\chi}^\mp \to \ell^+\ell^- + \slashed{E}_T.
\end{equation*}
Since there is no fatjet in the final states, our analysis is not directly applicable to this signal, but one can suitably modify the cuts to investigate it further. However, if a leptophobic $Z^\prime$ decays to a pair of neutralinos, those can decay to $W$ bosons and charginos, producing the lepton(s) plus fatjets signature. In that case, the signal will resemble ours, allowing recast, even though no RHN is involved in the process.

\section{Acknowledgements}
\noindent We thank T. Rizzo for insightful discussions on the $E_6$ embeddings. C. N. acknowledges the Department of Science and Technology (DST)-Inspire for his fellowship. We also thank Jai Bardhan for helping us speed up the input-feature generation pipeline and for helpful discussions.

\appendix

\begin{figure}[t!]
\centering
{\includegraphics[width=0.95\columnwidth]{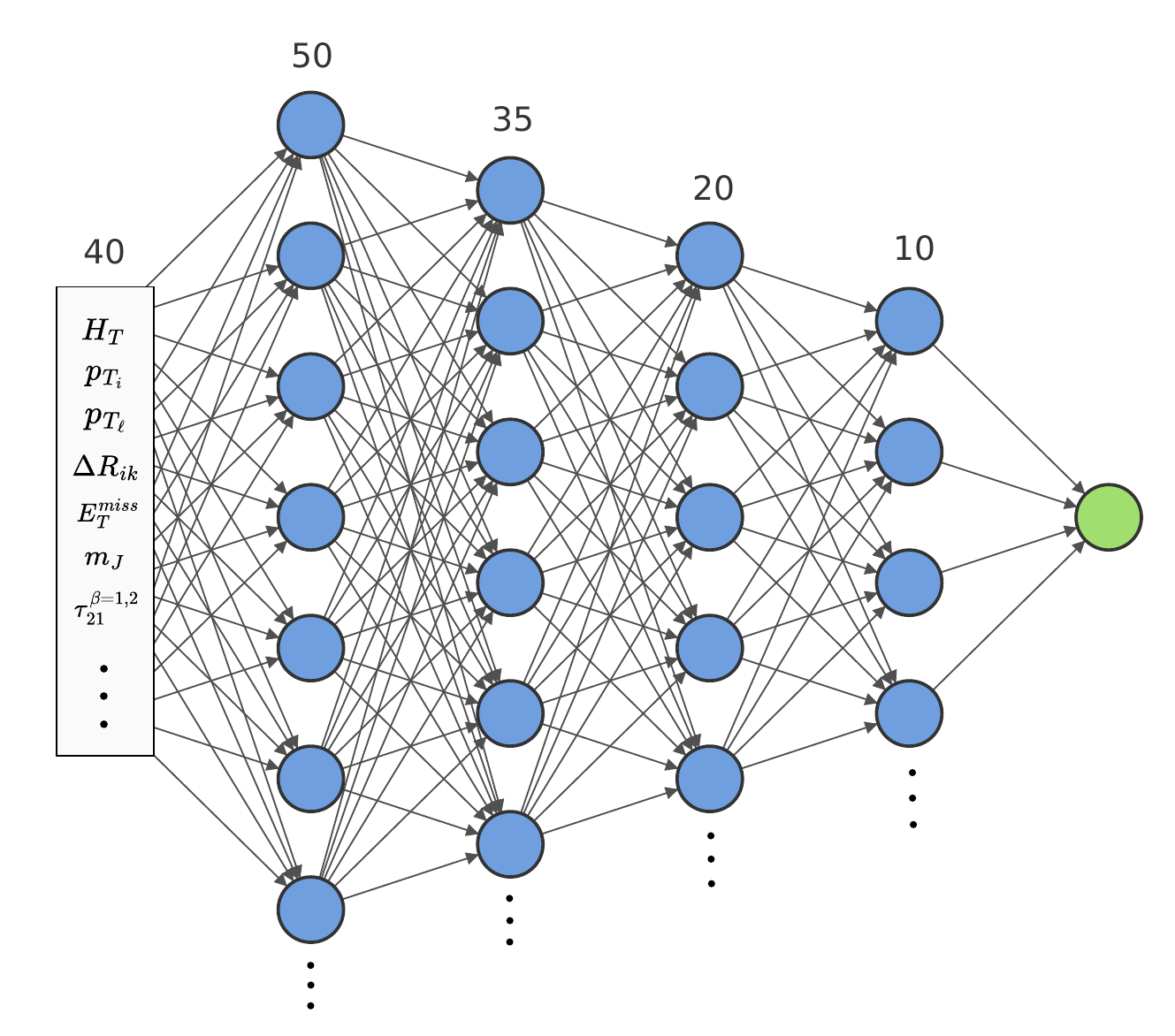}}\\
\caption{Schematic for the DNN model with four hidden layers. The number of nodes in a layer is shown on the top.}\label{fig:DNN}
\end{figure}
\section{Neural-network model} \label{App:DNN}
\noindent
A neural network is a series of perceptron blocks with a non-linear activation function. A perceptron is a linear transformation of the input vector of the form:
\begin{equation*}
    f(\textbf{x}) = \sigma (\textbf{W}\cdot \textbf{x} + \textbf{b}),
\end{equation*}
where $\sigma$ is an activation function. A DNN can effectively approximate any real continuous function provided the dimensions of hidden layers are sufficiently large. Here, we use a feed-forward DNN of four hidden layers with $50$, $35$, $20$, and $10$ nodes  respectively (see Fig.~\ref{fig:DNN}), with a ReLU activation function and batch normalisation~\cite{2015arXiv150203167I}. We implement it using the \texttt{TensorFlow} module~\cite{tensorflow2015-whitepaper} with the \texttt{AdamW} optimiser~\cite{2017arXiv171105101L} with a learning rate of $0.001$. The layers use  The final layer contains a single node with a sigmoid activation function, which outputs a value between $0$ and $1$. We set a cut on this output value while performing the classification task and estimate the signal and background efficiencies from the fraction of the events passing the threshold value. We incorporate dropouts~\cite{JMLR:v15:srivastava14a} with a probability of $0.1$ for training regularisation. For each mass point, we train the network for $200$ epochs with a batch size of $4096$. We use a basic binary cross-entropy loss while training. For tuning the network, a grid scan is performed for the benchmark point $(M_{Z'}, M_{N_R})$ = $(5.0, 2.0)$ TeV on the number of hidden layers, the number of nodes in each hidden layer and the learning rate.

\def\bibfont{\small}
\bibliography{ZprimeCollider}{}

\providecommand{\href}[2]{#2}\begingroup\raggedright\begin{thebibliography}{10}

\bibitem{Langacker:1980js}
P.~Langacker, \emph{{Grand Unified Theories and Proton Decay}},
  \href{http://dx.doi.org/10.1016/0370-1573(81)90059-4}{\emph{Phys. Rept.} {\bf
  72} (1981) 185}.

\bibitem{London:1986dk}
D.~London and J.~L. Rosner, \emph{{Extra Gauge Bosons in E(6)}},
  \href{http://dx.doi.org/10.1103/PhysRevD.34.1530}{\emph{Phys. Rev. D} {\bf
  34} (1986) 1530}.

\bibitem{Hewett:1988xc}
J.~L. Hewett and T.~G. Rizzo, \emph{{Low-Energy Phenomenology of Superstring
  Inspired E(6) Models}},
  \href{http://dx.doi.org/10.1016/0370-1573(89)90071-9}{\emph{Phys. Rept.} {\bf
  183} (1989) 193}.

\bibitem{Leike:1998wr}
A.~Leike, \emph{{The Phenomenology of extra neutral gauge bosons}},
  \href{http://dx.doi.org/10.1016/S0370-1573(98)00133-1}{\emph{Phys. Rept.}
  {\bf 317} (1999) 143--250}, [\href{http://arxiv.org/abs/hep-ph/9805494}{{\tt
  hep-ph/9805494}}].

\bibitem{Langacker:2008yv}
P.~Langacker, \emph{{The Physics of Heavy $Z^\prime$ Gauge Bosons}},
  \href{http://dx.doi.org/10.1103/RevModPhys.81.1199}{\emph{Rev. Mod. Phys.}
  {\bf 81} (2009) 1199--1228}, [\href{http://arxiv.org/abs/0801.1345}{{\tt
  0801.1345}}].

\bibitem{ATLAS:2019fgd}
{\bf ATLAS} collaboration, G.~Aad et~al., \emph{{Search for new resonances in
  mass distributions of jet pairs using 139 fb$^{-1}$ of $pp$ collisions at
  $\sqrt{s}=13$ TeV with the ATLAS detector}},
  \href{http://dx.doi.org/10.1007/JHEP03(2020)145}{\emph{JHEP} {\bf 03} (2020)
  145}, [\href{http://arxiv.org/abs/1910.08447}{{\tt 1910.08447}}].

\bibitem{CMS:2019gwf}
{\bf CMS} collaboration, A.~M. Sirunyan et~al., \emph{{Search for high mass
  dijet resonances with a new background prediction method in proton-proton
  collisions at $\sqrt{s} =$ 13 TeV}},
  \href{http://dx.doi.org/10.1007/JHEP05(2020)033}{\emph{JHEP} {\bf 05} (2020)
  033}, [\href{http://arxiv.org/abs/1911.03947}{{\tt 1911.03947}}].

\bibitem{ATLAS:2019erb}
{\bf ATLAS} collaboration, G.~Aad et~al., \emph{{Search for high-mass dilepton
  resonances using 139 fb$^{-1}$ of $pp$ collision data collected at
  $\sqrt{s}=$13 TeV with the ATLAS detector}},
  \href{http://dx.doi.org/10.1016/j.physletb.2019.07.016}{\emph{Phys. Lett. B}
  {\bf 796} (2019) 68--87}, [\href{http://arxiv.org/abs/1903.06248}{{\tt
  1903.06248}}].

\bibitem{CMS:2021ctt}
{\bf CMS} collaboration, A.~M. Sirunyan et~al., \emph{{Search for resonant and
  nonresonant new phenomena in high-mass dilepton final states at $ \sqrt{s} $
  = 13 TeV}}, \href{http://dx.doi.org/10.1007/JHEP07(2021)208}{\emph{JHEP} {\bf
  07} (2021) 208}, [\href{http://arxiv.org/abs/2103.02708}{{\tt 2103.02708}}].

\bibitem{ATLAS:2016gzy}
{\bf ATLAS} collaboration, M.~Aaboud et~al., \emph{{Search for resonances in
  diphoton events at $\sqrt{s}$=13 TeV with the ATLAS detector}},
  \href{http://dx.doi.org/10.1007/JHEP09(2016)001}{\emph{JHEP} {\bf 09} (2016)
  001}, [\href{http://arxiv.org/abs/1606.03833}{{\tt 1606.03833}}].

\bibitem{CMS:2016kgr}
{\bf CMS} collaboration, V.~Khachatryan et~al., \emph{{Search for high-mass
  diphoton resonances in proton\textendash{}proton collisions at 13 TeV and
  combination with 8 TeV search}},
  \href{http://dx.doi.org/10.1016/j.physletb.2017.01.027}{\emph{Phys. Lett. B}
  {\bf 767} (2017) 147--170}, [\href{http://arxiv.org/abs/1609.02507}{{\tt
  1609.02507}}].

\bibitem{ATLAS:2020fry}
{\bf ATLAS} collaboration, G.~Aad et~al., \emph{{Search for heavy diboson
  resonances in semileptonic final states in pp collisions at $\sqrt{s}=13$ TeV
  with the ATLAS detector}},
  \href{http://dx.doi.org/10.1140/epjc/s10052-020-08554-y}{\emph{Eur. Phys. J.
  C} {\bf 80} (2020) 1165}, [\href{http://arxiv.org/abs/2004.14636}{{\tt
  2004.14636}}].

\bibitem{CMS:2021klu}
{\bf CMS} collaboration, A.~Tumasyan et~al., \emph{{Search for heavy resonances
  decaying to WW, WZ, or WH boson pairs in the lepton plus merged jet final
  state in proton-proton collisions at $\sqrt{s}$ = 13 TeV}},
  \href{http://dx.doi.org/10.1103/PhysRevD.105.032008}{\emph{Phys. Rev. D} {\bf
  105} (2022) 032008}, [\href{http://arxiv.org/abs/2109.06055}{{\tt
  2109.06055}}].

\bibitem{CMS:2015fhb}
{\bf CMS} collaboration, V.~Khachatryan et~al., \emph{{Search for resonant $t
  \bar t$ production in proton-proton collisions at $\sqrt s=$ 8 TeV}},
  \href{http://dx.doi.org/10.1103/PhysRevD.93.012001}{\emph{Phys. Rev. D} {\bf
  93} (2016) 012001}, [\href{http://arxiv.org/abs/1506.03062}{{\tt
  1506.03062}}].

\bibitem{Ferrari:2002ac}
A.~Ferrari, \emph{{Study of the production of a new $Z^\prime$ boson and its
  decay into Majorana neutrinos in $p p$ collisions at $s$ = 14-TeV and in
  $e^{+} e^{-}$ collisions at $s$ = 3-TeV}},
  \href{http://dx.doi.org/10.1103/PhysRevD.65.093008}{\emph{Phys. Rev. D} {\bf
  65} (2002) 093008}.

\bibitem{Das:2017flq}
A.~Das, N.~Okada and D.~Raut, \emph{{Enhanced pair production of heavy Majorana
  neutrinos at the LHC}},
  \href{http://dx.doi.org/10.1103/PhysRevD.97.115023}{\emph{Phys. Rev. D} {\bf
  97} (2018) 115023}, [\href{http://arxiv.org/abs/1710.03377}{{\tt
  1710.03377}}].

\bibitem{Das:2017deo}
A.~Das, N.~Okada and D.~Raut, \emph{{Heavy Majorana neutrino pair productions
  at the LHC in minimal U(1) extended Standard Model}},
  \href{http://dx.doi.org/10.1140/epjc/s10052-018-6171-8}{\emph{Eur. Phys. J.
  C} {\bf 78} (2018) 696}, [\href{http://arxiv.org/abs/1711.09896}{{\tt
  1711.09896}}].

\bibitem{Cox:2017eme}
P.~Cox, C.~Han and T.~T. Yanagida, \emph{{LHC Search for Right-handed Neutrinos
  in $Z^\prime$ Models}},
  \href{http://dx.doi.org/10.1007/JHEP01(2018)037}{\emph{JHEP} {\bf 01} (2018)
  037}, [\href{http://arxiv.org/abs/1707.04532}{{\tt 1707.04532}}].

\bibitem{Chauhan:2021xus}
G.~Chauhan and P.~S.~B. Dev, \emph{{Interplay between resonant leptogenesis,
  neutrinoless double beta decay and collider signals in a model with flavor
  and CP symmetries}},
  \href{http://dx.doi.org/10.1016/j.nuclphysb.2022.116058}{\emph{Nucl. Phys. B}
  {\bf 986} (2023) 116058}, [\href{http://arxiv.org/abs/2112.09710}{{\tt
  2112.09710}}].

\bibitem{Das:2022rbl}
A.~Das, S.~Mandal, T.~Nomura and S.~Shil, \emph{{Heavy Majorana neutrino pair
  production from Z' at hadron and lepton colliders}},
  \href{http://dx.doi.org/10.1103/PhysRevD.105.095031}{\emph{Phys. Rev. D} {\bf
  105} (2022) 095031}, [\href{http://arxiv.org/abs/2202.13358}{{\tt
  2202.13358}}].

\bibitem{Ekstedt:2016wyi}
A.~Ekstedt, R.~Enberg, G.~Ingelman, J.~L\"ofgren and T.~Mandal,
  \emph{{Constraining minimal anomaly free $\mathrm{U}(1)$ extensions of the
  Standard Model}},
  \href{http://dx.doi.org/10.1007/JHEP11(2016)071}{\emph{JHEP} {\bf 11} (2016)
  071}, [\href{http://arxiv.org/abs/1605.04855}{{\tt 1605.04855}}].

\bibitem{Leontaris:1999wf}
G.~K. Leontaris and J.~Rizos, \emph{{New fermion mass textures from anomalous
  U(1) symmetries with baryon and lepton number conservation}},
  \href{http://dx.doi.org/10.1016/S0550-3213(99)00723-3}{\emph{Nucl. Phys. B}
  {\bf 567} (2000) 32--60}, [\href{http://arxiv.org/abs/hep-ph/9909206}{{\tt
  hep-ph/9909206}}].

\bibitem{Ekstedt:2017tbo}
A.~Ekstedt, R.~Enberg, G.~Ingelman, J.~L\"ofgren and T.~Mandal, \emph{{Minimal
  anomalous $\mathrm{U}(1)$ theories and collider phenomenology}},
  \href{http://dx.doi.org/10.1007/JHEP02(2018)152}{\emph{JHEP} {\bf 02} (2018)
  152}, [\href{http://arxiv.org/abs/1712.03410}{{\tt 1712.03410}}].

\bibitem{Arun:2022ecj}
M.~T. Arun, A.~Chatterjee, T.~Mandal, S.~Mitra, A.~Mukherjee and K.~Nivedita,
  \emph{{Search for the Z' boson decaying to a right-handed neutrino pair in
  leptophobic U(1) models}},
  \href{http://dx.doi.org/10.1103/PhysRevD.106.095035}{\emph{Phys. Rev. D} {\bf
  106} (2022) 095035}, [\href{http://arxiv.org/abs/2204.02949}{{\tt
  2204.02949}}].

\bibitem{Choudhury:2020cpm}
D.~Choudhury, K.~Deka, T.~Mandal and S.~Sadhukhan, \emph{{Neutrino and $Z'$
  phenomenology in an anomaly-free $\mathbf{U}(1)$ extension: role of
  higher-dimensional operators}},
  \href{http://dx.doi.org/10.1007/JHEP06(2020)111}{\emph{JHEP} {\bf 06} (2020)
  111}, [\href{http://arxiv.org/abs/2002.02349}{{\tt 2002.02349}}].

\bibitem{Deka:2021koh}
K.~Deka, T.~Mandal, A.~Mukherjee and S.~Sadhukhan, \emph{{Leptogenesis in an
  anomaly-free U(1) extension with higher-dimensional operators}},
  \href{http://dx.doi.org/10.1016/j.nuclphysb.2023.116213}{\emph{Nucl. Phys. B}
  {\bf 991} (2023) 116213}, [\href{http://arxiv.org/abs/2105.15088}{{\tt
  2105.15088}}].

\bibitem{ThomasArun:2021rwf}
M.~Thomas~Arun, T.~Mandal, S.~Mitra, A.~Mukherjee, L.~Priya and A.~Sampath,
  \emph{{Testing left-right symmetry with an inverse seesaw mechanism at the
  LHC}}, \href{http://dx.doi.org/10.1103/PhysRevD.105.115007}{\emph{Phys. Rev.
  D} {\bf 105} (2022) 115007}, [\href{http://arxiv.org/abs/2109.09585}{{\tt
  2109.09585}}].

\bibitem{Bhaskar:2023xkm}
A.~Bhaskar, Y.~Chaurasia, K.~Deka, T.~Mandal, S.~Mitra and A.~Mukherjee,
  \emph{{Right-handed neutrino pair production via second-generation
  leptoquarks}},
  \href{http://dx.doi.org/10.1016/j.physletb.2023.138039}{\emph{Phys. Lett. B}
  {\bf 843} (2023) 138039}, [\href{http://arxiv.org/abs/2301.11889}{{\tt
  2301.11889}}].

\bibitem{Mohapatra:1986aw}
R.~N. Mohapatra, \emph{{Mechanism for Understanding Small Neutrino Mass in
  Superstring Theories}},
  \href{http://dx.doi.org/10.1103/PhysRevLett.56.561}{\emph{Phys. Rev. Lett.}
  {\bf 56} (1986) 561--563}.

\bibitem{Mohapatra:1986bd}
R.~N. Mohapatra and J.~W.~F. Valle, \emph{{Neutrino Mass and Baryon Number
  Nonconservation in Superstring Models}},
  \href{http://dx.doi.org/10.1103/PhysRevD.34.1642}{\emph{Phys. Rev. D} {\bf
  34} (1986) 1642}.

\bibitem{Das:2017kkm}
D.~Das, K.~Ghosh, M.~Mitra and S.~Mondal, \emph{{Probing sterile neutrinos in
  the framework of inverse seesaw mechanism through leptoquark productions}},
  \href{http://dx.doi.org/10.1103/PhysRevD.97.015024}{\emph{Phys. Rev. D} {\bf
  97} (2018) 015024}, [\href{http://arxiv.org/abs/1708.06206}{{\tt
  1708.06206}}].

\bibitem{Jana:2019mez}
S.~Jana, P.~K. Vishnu and S.~Saad, \emph{{Minimal dirac neutrino mass models
  from $\hbox {U}(1)_{\mathrm{R}}$ gauge symmetry and left\textendash{}right
  asymmetry at colliders}},
  \href{http://dx.doi.org/10.1140/epjc/s10052-019-7441-9}{\emph{Eur. Phys. J.
  C} {\bf 79} (2019) 916}, [\href{http://arxiv.org/abs/1904.07407}{{\tt
  1904.07407}}].

\bibitem{Das:2019pua}
A.~Das, S.~Goswami, K.~N. Vishnudath and T.~Nomura, \emph{{Constraining a
  general U(1)$^\prime$ inverse seesaw model from vacuum stability, dark matter
  and collider}},
  \href{http://dx.doi.org/10.1103/PhysRevD.101.055026}{\emph{Phys. Rev. D} {\bf
  101} (2020) 055026}, [\href{http://arxiv.org/abs/1905.00201}{{\tt
  1905.00201}}].

\bibitem{Bandyopadhyay:2020djh}
P.~Bandyopadhyay, S.~Jangid and M.~Mitra, \emph{{Scrutinizing Vacuum Stability
  in IDM with Type-III Inverse seesaw}},
  \href{http://dx.doi.org/10.1007/JHEP02(2021)075}{\emph{JHEP} {\bf 02} (2021)
  075}, [\href{http://arxiv.org/abs/2008.11956}{{\tt 2008.11956}}].

\bibitem{Das:2021nqj}
A.~Das, S.~Goswami, V.~K.~N. and T.~K. Poddar, \emph{{Freeze-in sterile
  neutrino dark matter in a class of U$(1)^\prime$ models with inverse
  seesaw}},  \href{http://arxiv.org/abs/2104.13986}{{\tt 2104.13986}}.

\bibitem{Banerjee:2015gca}
S.~Banerjee, P.~S.~B. Dev, A.~Ibarra, T.~Mandal and M.~Mitra, \emph{{Prospects
  of Heavy Neutrino Searches at Future Lepton Colliders}},
  \href{http://dx.doi.org/10.1103/PhysRevD.92.075002}{\emph{Phys. Rev. D} {\bf
  92} (2015) 075002}, [\href{http://arxiv.org/abs/1503.05491}{{\tt
  1503.05491}}].

\bibitem{Chakraborty:2018khw}
S.~Chakraborty, M.~Mitra and S.~Shil, \emph{{Fat Jet Signature of a Heavy
  Neutrino at Lepton Collider}},
  \href{http://dx.doi.org/10.1103/PhysRevD.100.015012}{\emph{Phys. Rev. D} {\bf
  100} (2019) 015012}, [\href{http://arxiv.org/abs/1810.08970}{{\tt
  1810.08970}}].

\bibitem{Barducci:2022hll}
D.~Barducci and E.~Bertuzzo, \emph{{The see-saw portal at future Higgs
  factories: the role of dimension six operators}},
  \href{http://dx.doi.org/10.1007/JHEP06(2022)077}{\emph{JHEP} {\bf 06} (2022)
  077}, [\href{http://arxiv.org/abs/2201.11754}{{\tt 2201.11754}}].

\bibitem{Babu:1996vt}
K.~S. Babu, C.~F. Kolda and J.~March-Russell, \emph{{Leptophobic U(1) $s$ and
  the R($b$) - R($c$) crisis}},
  \href{http://dx.doi.org/10.1103/PhysRevD.54.4635}{\emph{Phys. Rev. D} {\bf
  54} (1996) 4635--4647}, [\href{http://arxiv.org/abs/hep-ph/9603212}{{\tt
  hep-ph/9603212}}].

\bibitem{Lopez:1996ta}
J.~L. Lopez and D.~V. Nanopoulos, \emph{{Leptophobic $Z^\prime$ in stringy
  flipped SU(5)}}, \href{http://dx.doi.org/10.1103/PhysRevD.55.397}{\emph{Phys.
  Rev. D} {\bf 55} (1997) 397--406},
  [\href{http://arxiv.org/abs/hep-ph/9605359}{{\tt hep-ph/9605359}}].

\bibitem{Rizzo:1998ut}
T.~G. Rizzo, \emph{{Gauge kinetic mixing and leptophobic $Z^\prime$ in E(6) and
  SO(10)}}, \href{http://dx.doi.org/10.1103/PhysRevD.59.015020}{\emph{Phys.
  Rev. D} {\bf 59} (1998) 015020},
  [\href{http://arxiv.org/abs/hep-ph/9806397}{{\tt hep-ph/9806397}}].

\bibitem{Leroux:2001fx}
K.~Leroux and D.~London, \emph{{Flavor changing neutral currents and
  leptophobic $Z^\prime$ gauge bosons}},
  \href{http://dx.doi.org/10.1016/S0370-2693(01)01489-7}{\emph{Phys. Lett. B}
  {\bf 526} (2002) 97--103}, [\href{http://arxiv.org/abs/hep-ph/0111246}{{\tt
  hep-ph/0111246}}].

\bibitem{Anastasopoulos:2006cz}
P.~Anastasopoulos, M.~Bianchi, E.~Dudas and E.~Kiritsis, \emph{{Anomalies,
  anomalous U(1)'s and generalized Chern-Simons terms}},
  \href{http://dx.doi.org/10.1088/1126-6708/2006/11/057}{\emph{JHEP} {\bf 11}
  (2006) 057}, [\href{http://arxiv.org/abs/hep-th/0605225}{{\tt
  hep-th/0605225}}].

\bibitem{Anastasopoulos:2008jt}
P.~Anastasopoulos, F.~Fucito, A.~Lionetto, G.~Pradisi, A.~Racioppi and Y.~S.
  Stanev, \emph{{Minimal Anomalous U(1)-prime Extension of the MSSM}},
  \href{http://dx.doi.org/10.1103/PhysRevD.78.085014}{\emph{Phys. Rev. D} {\bf
  78} (2008) 085014}, [\href{http://arxiv.org/abs/0804.1156}{{\tt 0804.1156}}].

\bibitem{Hettmansperger:2011bt}
H.~Hettmansperger, M.~Lindner and W.~Rodejohann, \emph{{Phenomenological
  Consequences of sub-leading Terms in See-Saw Formulas}},
  \href{http://dx.doi.org/10.1007/JHEP04(2011)123}{\emph{JHEP} {\bf 04} (2011)
  123}, [\href{http://arxiv.org/abs/1102.3432}{{\tt 1102.3432}}].

\bibitem{Eichten:1984eu}
E.~Eichten, I.~Hinchliffe, K.~D. Lane and C.~Quigg, \emph{{Super Collider
  Physics}}, \href{http://dx.doi.org/10.1103/RevModPhys.56.579}{\emph{Rev. Mod.
  Phys.} {\bf 56} (1984) 579--707}. [Addendum: Rev.Mod.Phys. 58, 1065--1073
  (1986)].

\bibitem{Alloul:2013bka}
A.~Alloul, N.~D. Christensen, C.~Degrande, C.~Duhr and B.~Fuks,
  \emph{{FeynRules 2.0 - A complete toolbox for tree-level phenomenology}},
  \href{http://dx.doi.org/10.1016/j.cpc.2014.04.012}{\emph{Comput. Phys.
  Commun.} {\bf 185} (2014) 2250--2300},
  [\href{http://arxiv.org/abs/1310.1921}{{\tt 1310.1921}}].

\bibitem{Alwall:2014hca}
J.~Alwall, R.~Frederix, S.~Frixione, V.~Hirschi, F.~Maltoni, O.~Mattelaer
  et~al., \emph{{The automated computation of tree-level and next-to-leading
  order differential cross sections, and their matching to parton shower
  simulations}}, \href{http://dx.doi.org/10.1007/JHEP07(2014)079}{\emph{JHEP}
  {\bf 07} (2014) 079}, [\href{http://arxiv.org/abs/1405.0301}{{\tt
  1405.0301}}].

\bibitem{Sjostrand:2014zea}
T.~Sj\"ostrand, S.~Ask, J.~R. Christiansen, R.~Corke, N.~Desai, P.~Ilten
  et~al., \emph{{An introduction to PYTHIA 8.2}},
  \href{http://dx.doi.org/10.1016/j.cpc.2015.01.024}{\emph{Comput. Phys.
  Commun.} {\bf 191} (2015) 159--177},
  [\href{http://arxiv.org/abs/1410.3012}{{\tt 1410.3012}}].

\bibitem{deFavereau:2013fsa}
{\bf DELPHES 3} collaboration, J.~de~Favereau, C.~Delaere, P.~Demin,
  A.~Giammanco, V.~Lema\^\i{}tre, A.~Mertens et~al., \emph{{DELPHES 3, A
  modular framework for fast simulation of a generic collider experiment}},
  \href{http://dx.doi.org/10.1007/JHEP02(2014)057}{\emph{JHEP} {\bf 02} (2014)
  057}, [\href{http://arxiv.org/abs/1307.6346}{{\tt 1307.6346}}].

\bibitem{Cacciari:2008gp}
M.~Cacciari, G.~P. Salam and G.~Soyez, \emph{{The anti-$k_t$ jet clustering
  algorithm}},
  \href{http://dx.doi.org/10.1088/1126-6708/2008/04/063}{\emph{JHEP} {\bf 04}
  (2008) 063}, [\href{http://arxiv.org/abs/0802.1189}{{\tt 0802.1189}}].

\bibitem{Thaler:2010tr}
J.~Thaler and K.~Van~Tilburg, \emph{{Identifying Boosted Objects with
  N-subjettiness}},
  \href{http://dx.doi.org/10.1007/JHEP03(2011)015}{\emph{JHEP} {\bf 03} (2011)
  015}, [\href{http://arxiv.org/abs/1011.2268}{{\tt 1011.2268}}].

\bibitem{Mitra:2016kov}
M.~Mitra, R.~Ruiz, D.~J. Scott and M.~Spannowsky, \emph{{Neutrino Jets from
  High-Mass $W_R$ Gauge Bosons in TeV-Scale Left-Right Symmetric Models}},
  \href{http://dx.doi.org/10.1103/PhysRevD.94.095016}{\emph{Phys. Rev. D} {\bf
  94} (2016) 095016}, [\href{http://arxiv.org/abs/1607.03504}{{\tt
  1607.03504}}].

\bibitem{delAguila:2008ir}
F.~del Aguila, S.~Bar-Shalom, A.~Soni and J.~Wudka, \emph{{Heavy Majorana
  Neutrinos in the Effective Lagrangian Description: Application to Hadron
  Colliders}},
  \href{http://dx.doi.org/10.1016/j.physletb.2008.11.031}{\emph{Phys. Lett. B}
  {\bf 670} (2009) 399--402}, [\href{http://arxiv.org/abs/0806.0876}{{\tt
  0806.0876}}].

\bibitem{Bhattacharya:2015vja}
S.~Bhattacharya and J.~Wudka, \emph{{Dimension-seven operators in the standard
  model with right handed neutrinos}},
  \href{http://dx.doi.org/10.1103/PhysRevD.94.055022}{\emph{Phys. Rev. D} {\bf
  94} (2016) 055022}, [\href{http://arxiv.org/abs/1505.05264}{{\tt
  1505.05264}}]. [Erratum: Phys.Rev.D 95, 039904 (2017)].

\bibitem{Liao:2016qyd}
Y.~Liao and X.-D. Ma, \emph{{Operators up to Dimension Seven in Standard Model
  Effective Field Theory Extended with Sterile Neutrinos}},
  \href{http://dx.doi.org/10.1103/PhysRevD.96.015012}{\emph{Phys. Rev. D} {\bf
  96} (2017) 015012}, [\href{http://arxiv.org/abs/1612.04527}{{\tt
  1612.04527}}].

\bibitem{Alcaide:2019pnf}
J.~Alcaide, S.~Banerjee, M.~Chala and A.~Titov, \emph{{Probes of the Standard
  Model effective field theory extended with a right-handed neutrino}},
  \href{http://dx.doi.org/10.1007/JHEP08(2019)031}{\emph{JHEP} {\bf 08} (2019)
  031}, [\href{http://arxiv.org/abs/1905.11375}{{\tt 1905.11375}}].

\bibitem{Mitra:2022nri}
M.~Mitra, S.~Mandal, R.~Padhan, A.~Sarkar and M.~Spannowsky, \emph{{Reexamining
  right-handed neutrino EFTs up to dimension six}},
  \href{http://dx.doi.org/10.1103/PhysRevD.106.113008}{\emph{Phys. Rev. D} {\bf
  106} (2022) 113008}, [\href{http://arxiv.org/abs/2210.12404}{{\tt
  2210.12404}}].

\bibitem{Langacker:1998tc}
P.~Langacker and J.~Wang, \emph{{U(1)-prime symmetry breaking in supersymmetric
  E(6) models}},
  \href{http://dx.doi.org/10.1103/PhysRevD.58.115010}{\emph{Phys. Rev. D} {\bf
  58} (1998) 115010}, [\href{http://arxiv.org/abs/hep-ph/9804428}{{\tt
  hep-ph/9804428}}].

\bibitem{Araz:2017wbp}
J.~Y. Araz, G.~Corcella, M.~Frank and B.~Fuks, \emph{{Loopholes in $Z^\prime$
  searches at the LHC: exploring supersymmetric and leptophobic scenarios}},
  \href{http://dx.doi.org/10.1007/JHEP02(2018)092}{\emph{JHEP} {\bf 02} (2018)
  092}, [\href{http://arxiv.org/abs/1711.06302}{{\tt 1711.06302}}].

\bibitem{Frank:2020kvp}
M.~Frank, Y.~Hi\c{c}y\i{}lmaz, S.~Moretti and O.~\"Ozdal, \emph{{Leptophobic
  $Z^\prime$ bosons in the secluded UMSSM}},
  \href{http://dx.doi.org/10.1103/PhysRevD.102.115025}{\emph{Phys. Rev. D} {\bf
  102} (2020) 115025}, [\href{http://arxiv.org/abs/2005.08472}{{\tt
  2005.08472}}].

\bibitem{2015arXiv150203167I}
S.~{Ioffe} and C.~{Szegedy}, \emph{{Batch Normalization: Accelerating Deep
  Network Training by Reducing Internal Covariate Shift}},
  \href{http://arxiv.org/abs/1502.03167}{{\tt 1502.03167}}.

\bibitem{tensorflow2015-whitepaper}
M.~Abadi, A.~Agarwal, P.~Barham, E.~Brevdo, Z.~Chen, C.~Citro et~al.,
  \emph{{TensorFlow}: Large-scale machine learning on heterogeneous systems},
  \href{http://arxiv.org/abs/1603.04467}{{\tt 1603.04467}}. Software available
  from \href{https://www.tensorflow.org/}{tensorflow.org}.

\bibitem{2017arXiv171105101L}
I.~{Loshchilov} and F.~{Hutter}, \emph{{Decoupled Weight Decay
  Regularization}},
  {\emph{\href{https://openreview.net/forum?id=Bkg6RiCqY7}{International
  Conference on Learning Representations}} (2019) },
  [\href{http://arxiv.org/abs/1711.05101}{{\tt 1711.05101}}].

\bibitem{JMLR:v15:srivastava14a}
N.~Srivastava, G.~Hinton, A.~Krizhevsky, I.~Sutskever and R.~Salakhutdinov,
  \emph{Dropout: A simple way to prevent neural networks from overfitting},
  {\emph{Journal of Machine Learning Research} {\bf 15} (2014) 1929--1958}.

\end{thebibliography}\endgroup
\bibliographystyle{JHEPCust}
\end{document}